\documentclass[%
reprint,
amsmath,amssymb,
]{revtex4-2}

\usepackage{epsfig}
\usepackage{dcolumn}
\usepackage{bm}
\usepackage{graphicx}
\usepackage{color}

\begin{document}

\title{Kinetic scales dominated by magnetic helicity in space plasmas} 
\author{A. Bershadskii}
\affiliation{ICAR, P.O. Box 31155, Jerusalem 91000, Israel}
\email{bershads@gmail.com}

\begin{abstract}\
It is shown, using results of fully kinetic 3D numerical simulations and observations in solar wind and Earth's magnetosphere that the transition from deterministic chaos to turbulence at kinetic (sub-ion) scales in space plasmas is generally dominated by magnetic helicity (an adiabatic invariant in a weakly dissipative plasma) directly or through the Kolmogorov-Iroshnikov phenomenology (the magneto-inertial range of scales as a precursor of hard turbulence). The magneto-inertial range of scales at sub-electron scales has been also briefly discussed. Despite the considerable differences in the scales and physical parameters, the results of numerical simulations are in quantitative agreement with the space observations in the frames of this approach.
\end{abstract}

\maketitle
\section{Introduction}
 The space plasmas (in the solar wind and planet's magnetospheres) are collisionless and it can be expected that kinetic scales play a special role in their dynamics. A vigorous discussion of the physical processes at these scales takes place in the literature. On the one hand, the kinetic scales are suggested as the dissipative ones on the other hand existence of a second inertial range at these scales is also expected. Laboratory experiments cannot help to resolve these problems. Therefore, one should use numerical simulations and measurements directly produced in the space plasmas. Considerable advances in these directions are related to the fast development of computers and the measuring technology onboard the special spacecraft. However, differences in the physical parameters available in the present numerical simulations and those characteristic to the space plasmas still questioned the applicability of the results obtained in the numerical simulation to the space plasmas. On the other hand, the numerous measurements directly produced in the space plasmas show great variability of the physical processes. The absence of a universal theoretical or even phenomenological approach to these chaotic/turbulent processes makes the problem rather difficult. \\

    The notion of smoothness can be useful for the classification of chaotic/turbulent dynamical regimes. Then the spectral analyses can be functional for this purpose. The  stretched exponential power spectra are typical for the smooth dynamics
    
\begin{equation}
E(k) \propto \exp-(k/k_{\beta})^{\beta} 
\end{equation}   

 here $k$ is the wavenumer and $1 \geq \beta > 0$. Deterministic chaos is typically characterized by the particular value $\beta =1$ (see, for instance, \cite{fm, swinney1, mm1, mm2, mm3, kds} and references therein)
\begin{equation}
E(k) \propto \exp(-k/k_c).  
\end{equation}

 For $1 > \beta$ the chaotic dynamics is still smooth but already not deterministic (this phenomenon will be called distributed chaos and a clarification of the term will be provided below). The term ``soft turbulence'' introduced in paper \cite{wu} can be also appropriate.\\
 
  For non-smooth dynamics (hard turbulence \cite{wu}) the power-law (scaling) spectra are typical. \\

\begin{figure} \vspace{-0.2cm}\centering \hspace{-1.1cm}
\epsfig{width=.45\textwidth,file=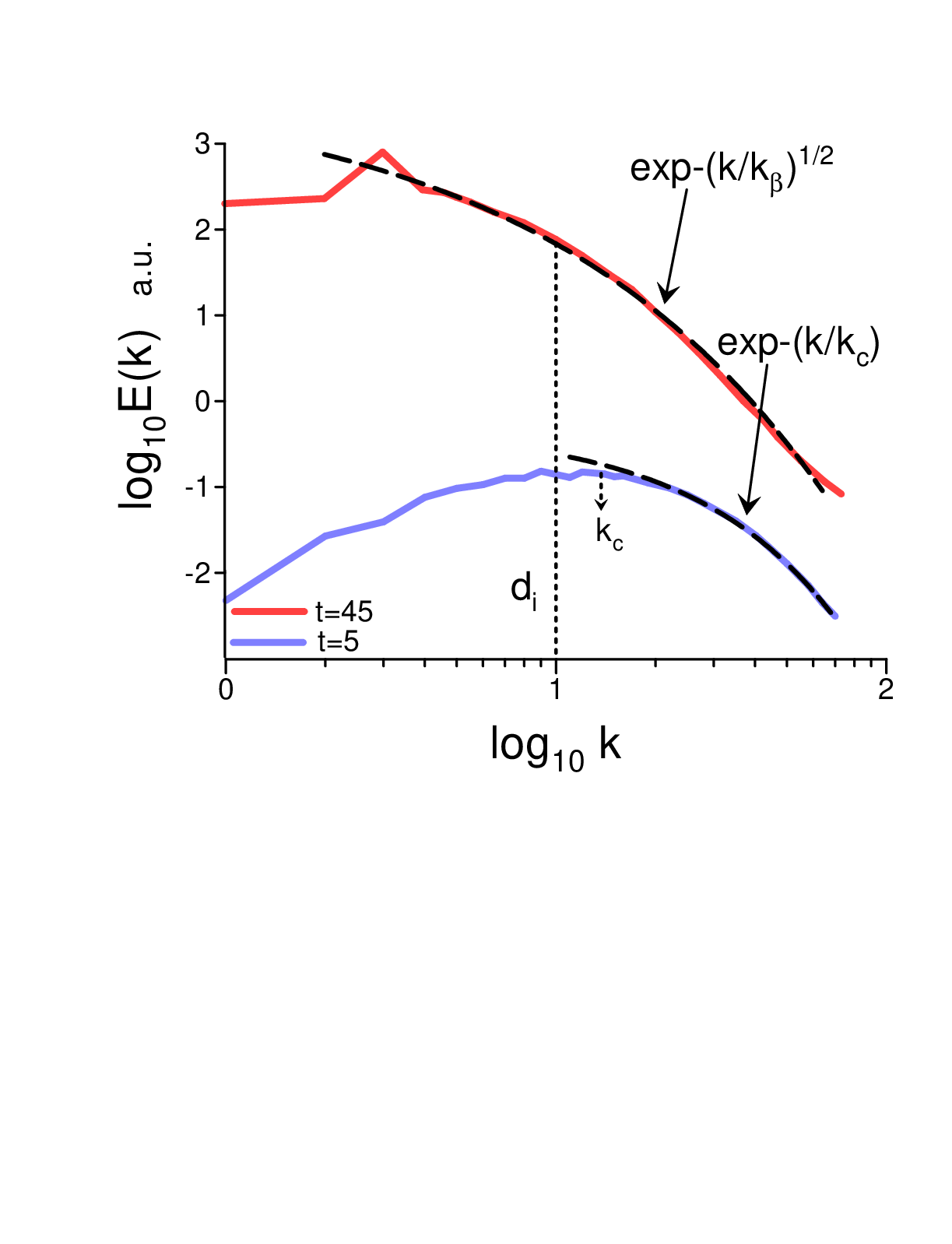} \vspace{-4.2cm}
\caption{Magnetic energy spectra generated by a Hall MHD dynamo at different times of the dynamo evolution: $t=5$ corresponds to an earlier stage and $t=45$ corresponds to the saturation stage of the small-scale Hall MHD dynamo (direct numerical simulations). The spectra are vertically shifted for clarity.} 
\end{figure}

 Figure 1, for instance, shows the evolution of the magnetic energy spectrum generated by small-scale Hall MHD dynamo from the deterministic chaos at an earlier stage for $t=5$ (in terms of the direct numerical simulation) to the distributed chaos with $\beta =1/2$ at a saturated stage for $t=45$. The spectral data were taken from Fig. 2b  of a paper \cite{map}, where results of a direct numerical simulation were reported.\\ 
  
  In the ideal Hall magnetohydrodynamics ions are not tied to the magnetic field due to the ions' inertia. Electrons (which are tied to the magnetic field) and the ions are assumed to decouple at the ions' inertial length $d_i$ (indicated in Fig. 1 using the vertical dotted line). \\
  
   The equations of the Hall incompressible magnetohydrodynamics can be written as
\begin{equation}
\partial_t {\bf u} + ({\bf u}\cdot \nabla) {\bf u} =  - \nabla P + \left({\bf b}\cdot \nabla\right){\bf b} +\nu\nabla^2 {\bf u} 
\end{equation}
\begin{equation}
\partial_t {\bf b} = - \left({\bf u}\cdot {\bf \nabla}\right){\bf b} + \left({\bf b}\cdot \nabla\right){\bf u} + \eta\nabla^2 \bf{b} +\nonumber
\end{equation}
\begin{equation}
 + d_i\left({\bf j}\cdot {\bf \nabla}\right){\bf b} -d_i\left({\bf b}\cdot \nabla\right){\bf j} 
\end{equation}
\begin{equation}
\nabla \cdot {\bf u} = 0, ~~~   {\bf \nabla} \cdot {\bf b} =  0, 
\end{equation}
 where ${\bf u}$ and ${\bf b}$ are the velocity and magnetic fields respectively, $P = p +{\bf b}^2/2$, $p$ is the pressure field, ${\bf j}= \nabla \times {\bf b}$. The Hall effect was taken into account by the addition of the last two terms into Eq. (4) . \\

The dashed curves in Fig. 1 indicate the best fit by the exponential spectral law Eq. (2) (deterministic chaos) at $t=5$ and by the stretched exponential spectral law Eq. (1) with $\beta =2$ (distributed chaos) at $t=45$. \\
    
    It will be shown in the next Section II that the value of $\beta = 1/2$ corresponds to the distributed chaos dominated by the magnetic helicity.  Then in Section III a magneto-inertial range of scales dominated by magnetic heilicity will be introduced in the frames of the Kolmogorov-Iroshnikov phenomenology (which is characterized by different values of $\beta$, see also Ref. \cite{ber4}). In Section IV the theoretical considerations will be compared to results of fully kinetic 3D numerical simulations. In Section V the comparison will be extended to the results of direct measurements in the solar wind and Earth's magnetosphere.
   
\section{Distributed chaos and magnetic helicity}  
 
 \subsection{Magnetic helicity}   
 
 The ideal magnetohydrodynamics has three quadratic (fundamental) invariants: total energy, cross and magnetic helicity \cite{mt}.  The total energy and magnetic helicity are ideal invariants for the Hall magnetohydrodynamics as well \cite{pm,sch1}. \\
  
  The magnetic helicity is
\begin{equation}
 h_m = \langle {\bf a} {\bf b} \rangle   
\end{equation}
where the fluctuating magnetic field is ${\bf b} = [{\nabla \times \bf a}]$, and a spatial average is denoted as $\langle ... \rangle$. The fluctuating magnetic potential ${\bf a}$ and ${\bf b}$ have zero means, and $\nabla \cdot {\bf a} =0$. \\

 A uniform magnetic field ${\bf B_0}$ generally breaks the invariance of magnetic helicity. A generalized magnetic helicity was introduced in the paper \cite{mg} in the form
\begin{equation}
 \hat{h} = h_m + 2{\bf B_0}\cdot \langle {\bf A}  \rangle 
 \end{equation}
here ${\bf B} = {\bf B_0} + {\bf b}$, ${\bf A} = {\bf A_0} +{\bf a}$, and ${\bf b} = [{\nabla \times \bf a}]$. It can be shown that in the ideal magnetohydrodynamics \cite{mg} (see also Ref. \cite{shebalin})
\begin{equation}
 \frac{d \hat{h}}{d t} =  0, 
\end{equation}
i.e. the generalized magnetic helicity  $\hat{h} $ is an ideal invariant in the presence of a uniform magnetic field.\\

\subsection{Distributed chaos in magnetic field dynamics driven by magnetic helicity} 
   
     As we can see in the Introduction (Fig. 1) temporal evolution of the magnetic field generated by the small-scale Hall MHD dynamo results in a transition from deterministic spatial chaos at the beginning of the dynamo process ($\beta =1$ at $t=5$) to a spatial distributed chaos ($\beta =1/2$) at a saturation stage (t=45).
     
     Let us understand the transition. The evolution of the dynamo-generated magnetic field can result in the random fluctuations of the characteristic scale $k_c$ in Eq. (2). To take this phenomenon into account we  should use an ensemble averaging 
\begin{equation}
E(k) \propto \int_0^{\infty} P(k_c) \exp -(k/k_c)dk_c 
\end{equation}  
where a probability {\it distribution} $P(k_c)$ characterizes the random fluctuations of the characteristic scale $k_c$. This is the reason why the smooth non-deterministic chaotic dynamics was called `distributed chaos'.\\

  Let us use the dimensional considerations and a scaling relationship
\begin{equation}
B_c \propto |h_m|^{1/2} k_c^{1/2}   
\end{equation}
to find the probability distribution $P(k_c)$ for magnetic field dynamics driven by the magnetic helicity (here $B_c$ is a characteristic value of the magnetic field). \\

   Indeed, let the characteristic value of the magnetic field $B_c$ be half-normally distributed $P(B_c) \propto \exp- (B_c^2/2\sigma^2)$ \cite{my} (it is a normal distribution with zero mean truncated to only have nonzero probability density for positive values of its argument: if $B$ is a normal random variable,  then $B_c = |B|$ has a half-normal distribution \cite{jkb}). Then from Eq. (10) we obtain that the characteristic value of the wavenumber $k_c$ has the chi-squared ($\chi^{2}$) distribution
\begin{equation}
P(k_c) \propto k_c^{-1/2} \exp-(k_c/4k_{\beta})  
\end{equation}
where $k_{\beta}$ is a new constant. 

   Substituting Eq. (11) into Eq. (9) we obtain
\begin{equation}
E(k) \propto \exp-(k/k_{\beta})^{1/2}  
\end{equation}

   It should be noted that $h_m$ was used in the scaling estimate (10) as an (ideal) invariant. It is well known, however, that in weakly non-dissipative case the magnetic helicity $h_m$ is also almost conserved (because there are no physical processes that can effectively transfer the magnetic helicity to the resistive scales) while the magnetic and total energy can efficiently dissipate. It means that $h_m$ is an {\it adiabatic} invariant in weakly non-dissipative case.\\ 
  
\subsection{Spontaneous breaking of local reflectional symmetry}

  In a special case of net (global) reflectional symmetry, the mean (global) magnetic helicity is equal to zero, whereas the point-wise magnetic helicity is not (due to the spontaneous breaking of the local reflectional symmetry). This can be considered as a built-in property of chaotic/turbulent flows. The appearance of the blobs with non-zero magnetic/kinetic helicity should accompany the spontaneous symmetry breaking \cite{mt,moff1,moff2,kerr,hk,lt}. The magnetic blobs are bounded by the magnetic surfaces with ${\bf b_n}\cdot {\bf n}=0$ (where ${\bf n}$ is a unit normal to the boundary of the blob). Sign-defined magnetic helicity of each magnetic blob is an ideal (adiabatic, see above) invariant \cite{mt}. These magnetic blobs could be numbered with $H_j^{\pm}$ denoting the helicity of the blob with number $j$ and the blob's helicity sign (`+'  or `-'):
\begin{equation}
 H_j^{\pm} = \int_{V_j} ({\bf a} ({\bf x},t) \cdot  {\bf b} ({\bf x},t)) ~ d{\bf x} 
\end{equation}  
  
  Then one can consider the adiabatic invariant 
\begin{equation}
{\rm I^{\pm}} = \lim_{V \rightarrow  \infty} \frac{1}{V} \sum_j H_{j}^{\pm}   
\end{equation}
 The summation is over the blobs with a certain sign only (`+' or `-'), and $V$ is the total volume of the blobs over which the summation was made.  \\

\begin{figure} \vspace{-0.7cm}\centering
\epsfig{width=.46\textwidth,file=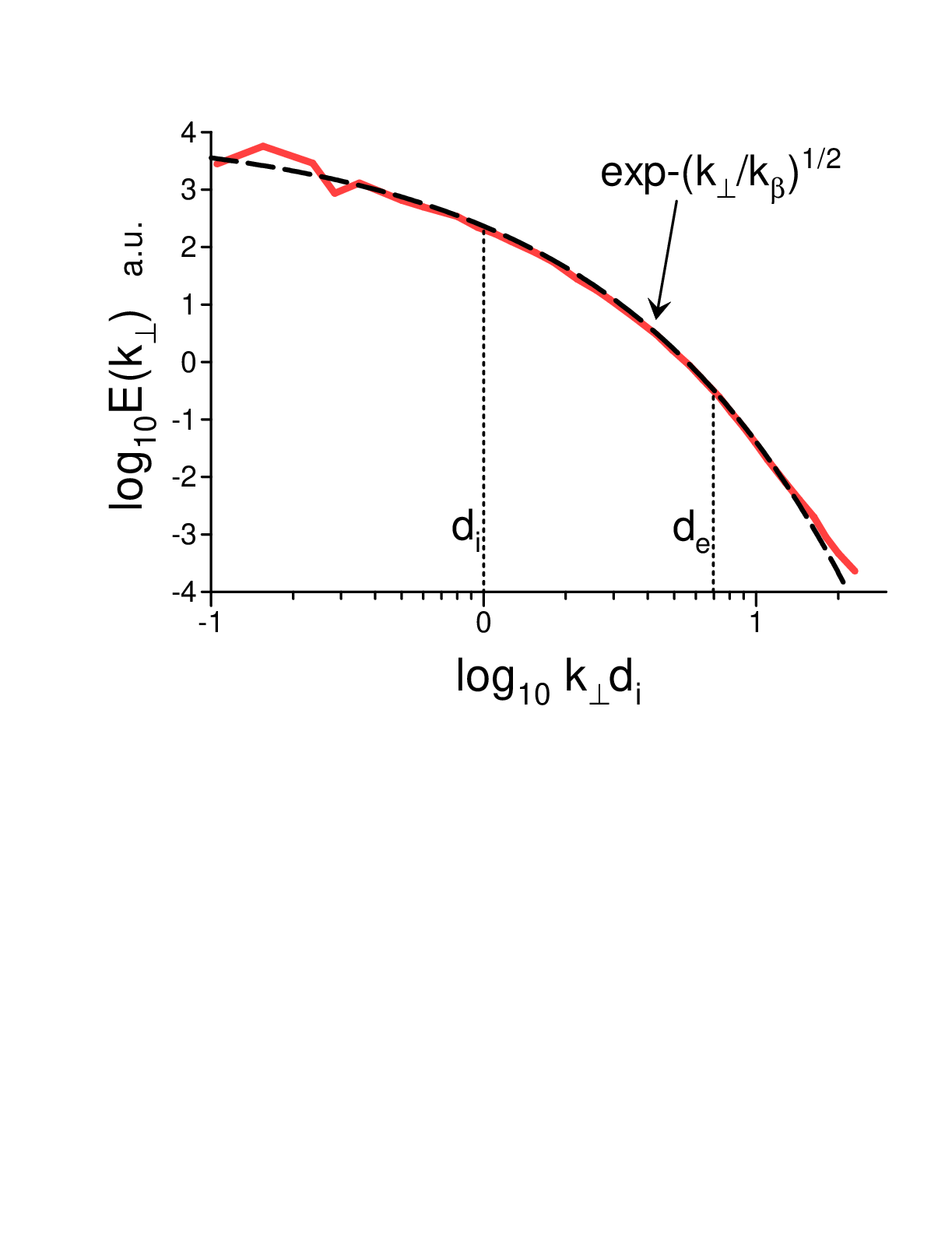} \vspace{-4.8cm}
\caption{Magnetic energy spectrum vs. $k_{\perp}d_i$ obtained in a recent 3D fully kinetic (Vlasov–Maxwell equations) particle-in-cell simulation with a uniform guide magnetic field ${\bf B}_0$.} 
\end{figure}

  For the case of the local reflectional symmetry breaking the adiabatic invariant ${\rm I^{\pm}}$ Eq. (14) could be used instead of $h_m$ for the case of the local symmetry breaking
\begin{equation}
B_c \propto |{\rm I^{\pm}}|^{1/2} k_c^{1/2}   
\end{equation}
and the spectrum Eq. (12) can be obtained for this case well.\\

  In the case of the non-zero mean magnetic field ${\bf B}_0$ the magnetic helicity $h_m$ can be replaced by the generalized magnetic helicity $\hat{h}$ Eq. (7). Since the $\hat{h}$ has the same dimensionality as the $h_m$ magnetic energy spectrum will be the same Eq. (12) (analogously for the case of the local reflectional symmetry breaking).\\
 
  In Fig. 1 we have already seen the stretched exponential spectrum Eq. (12) obtained at a saturation stage of the small-scale Hall MHD dynamo. \\
  
    Figure 2 shows the magnetic energy spectrum vs. $k_{\perp} =(k_x^2+k_y^2)^{1/2}$ obtained in a recent 3D fully kinetic (Vlasov–Maxwell equations) particle-in-cell simulation with a uniform guide magnetic field ${\bf B}_0$ along the $z$-axis. The spectral data were taken from Fig. 1c of the paper \cite{cs}. The electron skin depth $d_e = c/\omega_{pe}$ where $\omega_{pe}= (4\pi n_0 e^2/m_e)^{1/2}$ is the electron plasma frequency and $2n_0$ is the total particle density. \\
  
    The dashed curve in Fig. 2 indicates the best fit by the stretched exponential spectral law Eq. (12) (distributed chaos dominated by magnetic helicity).  The dotted vertical lines indicate the positions of the ions' inertial length $d_i$ and electron skin depth $d_e$.\\
    
    One can see that Eq. (12) can well describe not only the sub-ion range but also a certain part of the sub-electron range of the wavenumbers.

 \section{Magneto-inertial range of scales}   

    For the high Reynolds numbers an inertial range of scales is expected in hydrodynamic turbulence. In this range, the energy spectra depend on the energy dissipation rate $\varepsilon$ only \cite{my}. 
    
    A magneto-inertial range of scales can be introduced in magnetohydrodynamics and its modifications related to the kinetic scales \cite{ber4}.   Two parameters: the total energy dissipation rate $\varepsilon$ and the magnetic helicity dissipation rate $\varepsilon_h$ (or dissipation rate of its modification $I^{\pm}$ Eq. (14)) control the magnetic field dynamics in the magneto-inertial range of scales. 
    
    This approach is analogous to that with the passive scalar in the turbulent flows (the Corrsin-Obukhov approach) where two controlling parameters: the energy dissipation rate and the passive scalar dissipation rate, determined the inertial-convective range \cite{my} (see also Ref. \cite{bs1}). \\
    
    Following this analogy one can replace the estimates Eqs. (10) and (15) by the estimate
\begin{equation}
 B_c \propto \varepsilon_h^{1/2} ~\varepsilon^{-1/6}~k_c^{1/6}  
\end{equation}
 for the magneto-inertial range dominated by magnetic helicity.\\
 
  In the presence of a considerable mean magnetic field ${\bf B}_0$ the energy dissipation rate $\varepsilon$ in Eq. (16) can be replaced by a more appropriate $(\varepsilon \widetilde{B}_0)$ (here  $\widetilde{B}_0 = B_0/\sqrt{\mu_0\rho}$ is the normalized mean magnetic field, having the same dimension as velocity ) \cite{ir}. The appropriate dimensional considerations result in
\begin{equation}
 B_c \propto \varepsilon_{\hat{h}}^{1/2}~ (\varepsilon \widetilde{B}_0)^{-1/8}  k_c^{1/8} 
\end{equation}
where $\varepsilon_{\hat{h}}$ is the modified magnetic helicity dissipation rate (see Eq. (7)). \\
 
 The above-considered estimates Eq. (10), (15), (16), and (17) can be generalized as
\begin{equation}
 B_c \propto k_c^{\alpha}   
\end{equation}
  
  The stretched exponential form of the distributed chaos spectra 
\begin{equation}
 \int_0^{\infty}  P(k_c) \exp -(k/k_c)dk_c \propto \exp-(k/k_{\beta})^{\beta} 
\end{equation}  
can be used for finding the distribution $P(k_c)$ at large $k_c$ \cite{jon}
\begin{equation}
P(k_c) \propto k_c^{-1 + \beta/[2(1-\beta)]}~\exp(-\gamma k_c^{\beta/(1-\beta)}) 
\end{equation}   
  
   Then, a relationship between the $\beta$ and $\alpha$ can be derived (using some algebra) from the Eqs. (18) and (20) for the half-normally distributed $B_c$ 
\begin{equation}
\beta = \frac{2\alpha}{1+2\alpha}  
\end{equation}

  It follows from Eq. (21) that for $\alpha =1/6$ (see Eq. (16)) 
\begin{equation}
 E(k) \propto \exp-(k/k_{\beta})^{1/4}  
\end{equation}
 whereas for  $\alpha =1/8$ (see Eq. (17))
\begin{equation}
 E(k) \propto \exp-(k/k_{\beta})^{1/5}  
\end{equation}
   These spectra can be considered as a precursor of hard turbulence (see Introduction). \\
 
\begin{figure} \vspace{-0.9cm}\centering 
\epsfig{width=.42\textwidth,file=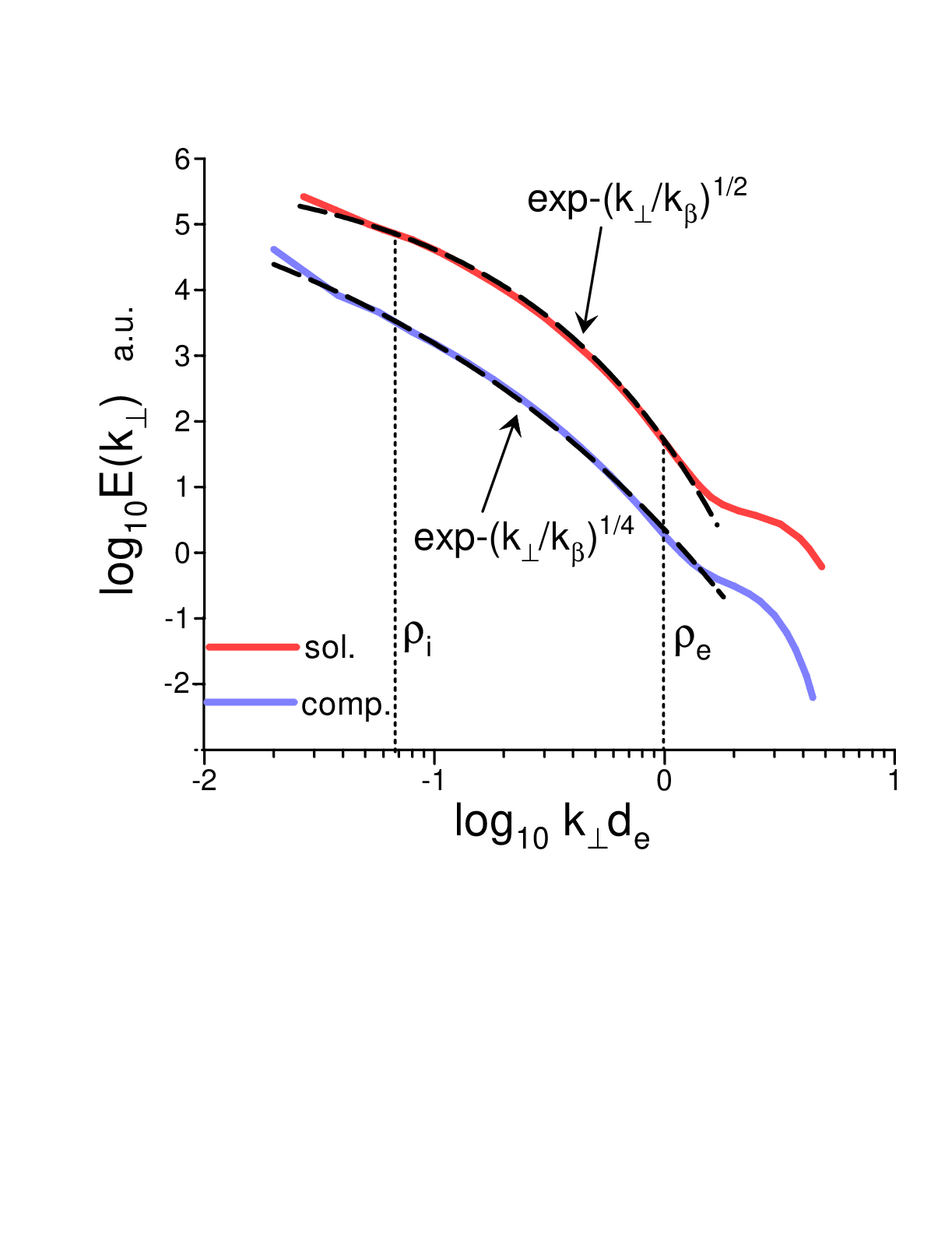} \vspace{-3.3cm}
\caption{Magnetic energy spectra obtained in 3D fully kinetic particle-in-cell semirelativistic simulations with the solenoidal (top curve) and compressible (bottom curve) body driving.} 
\end{figure}

 \section{Magneto-inertial range in fully kinetic 3D numerical simulations} 

    Figure 3 shows magnetic energy spectra obtained in recent 3D fully kinetic particle-in-cell semirelativistic numerical simulations with the solenoidal (top curve) and compressible (bottom curve) body driving in the presence of a uniform guide magnetic field ${\bf B}_0$ along the $z$-axis. The spectral data were taken from a recent paper \cite{zd}. In the semirelativistic regime, ions are still subrelativistic while electrons are already relativistically hot. The semirelativistic regime can be relevant for the hot accretion 
flows. The vertical dotted lines indicate the positions of the $\rho_i$ and $\rho_e$ which denote the ion and electron gyroradii respectively. \\

   The dashed curves in Fig. 3 indicate the best fit by the stretched exponential spectral law Eq. (12) (distributed chaos dominated by magnetic helicity for the solenoidal driving) and Eq. (22)   (distributed chaos in the magneto-inertial range for the compressible driving). It should be noted that for compressible driving Eq. (22) (and not Eq. (23)) was more appropriate despite the presence of the mean magnetic field (that may be related to a diminished role of the Alfv\'{e}n waves in this case, cf ordinary magnetohydrodynamics \cite{ber4}). \\
 
    This observation can be supported by Fig. 4 which shows a perpendicular magnetic energy spectrum obtained in 3D fully kinetic (the Vlasov–Maxwell equations) particle-in-cell relativistic numerical simulation. The spectral data were taken from Fig. 2 of the paper \cite{wan}. In this case a uniform guide magnetic field ${\bf B}_0$ was also present. 
    The dashed curve indicates the stretched exponential spectrum Eq. (22) (cf Fig. 3).\\
    
\begin{figure} \vspace{-1cm}\centering \hspace{-1cm}
\epsfig{width=.45\textwidth,file=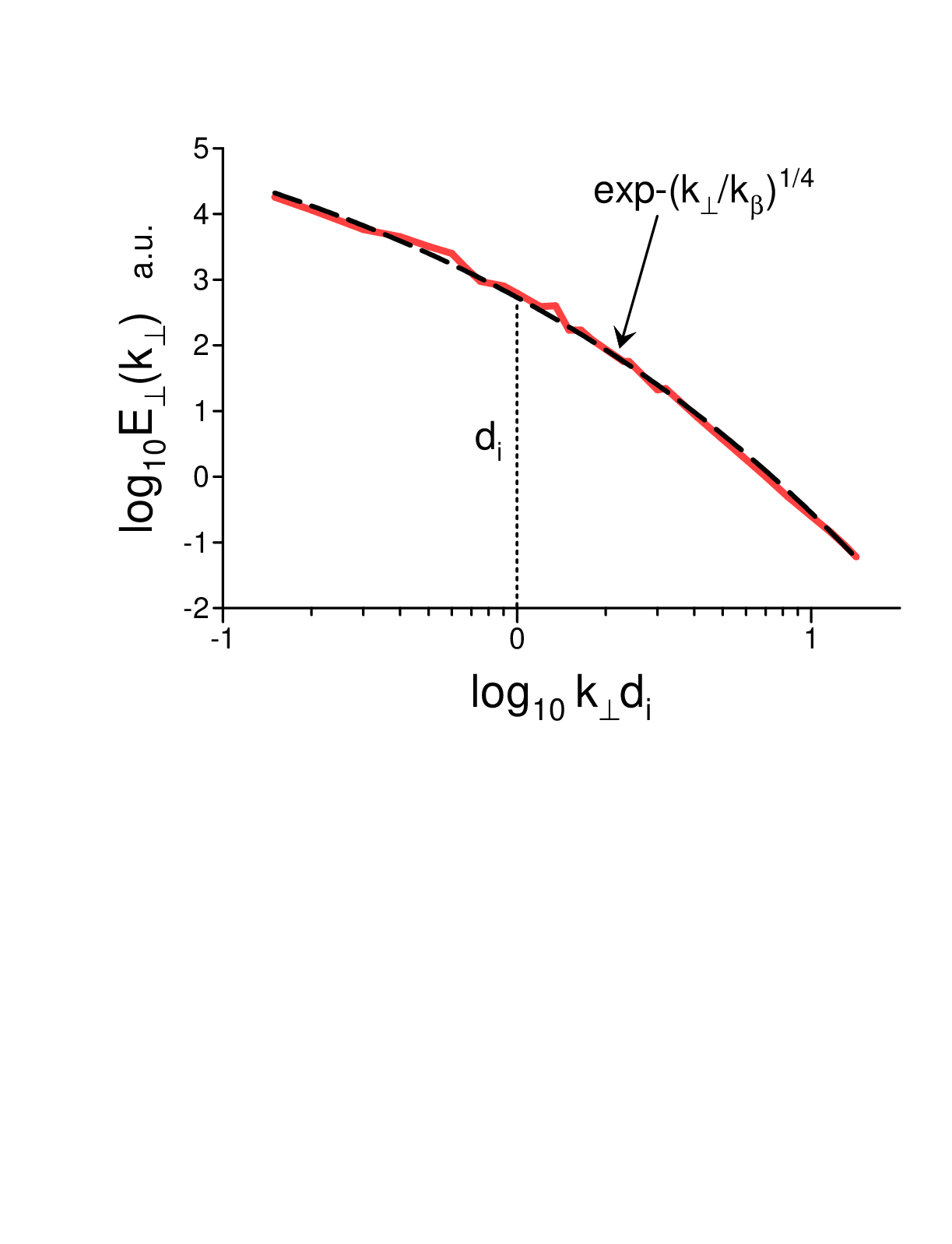} \vspace{-4.5cm}
\caption{Magnetic energy spectra obtained in a 3D fully kinetic particle-in-cell relativistic simulation.} 
\end{figure}

     In a paper \cite{thd} the gyroaveraged Vlasov–Maxwell equations for non-relativistic electrons and protons were numerically solved. The equilibrium distribution functions for electrons and protons were taken Maxwellian. A computational spatial periodic domain was elongated along a uniform (equilibrium) imposed magnetic field ${\bf B}_0$. 
     The simulation was driven by a Langevin antenna which was coupled to the parallel vector potential \cite{th}.\\

   Figure 5 shows the magnetic energy spectrum perpendicular to the uniform magnetic field ${\bf B}_0$ vs $k_{\perp} \rho_i$ ($\rho_e$ and $\rho_i$ are electron and ion gyroradii respectively). The spectral data were taken from Fig. 1 of the Ref. \cite{thd}. 
   
   The dashed curve indicates the stretched exponential spectrum Eq. (23). The dashed vertical lines indicate the positions of the gyroradii $\rho_e$ and $\rho_i$.\\
 
\begin{figure} \vspace{-0.5cm}\centering \hspace{-1.3cm}
\epsfig{width=.47\textwidth,file=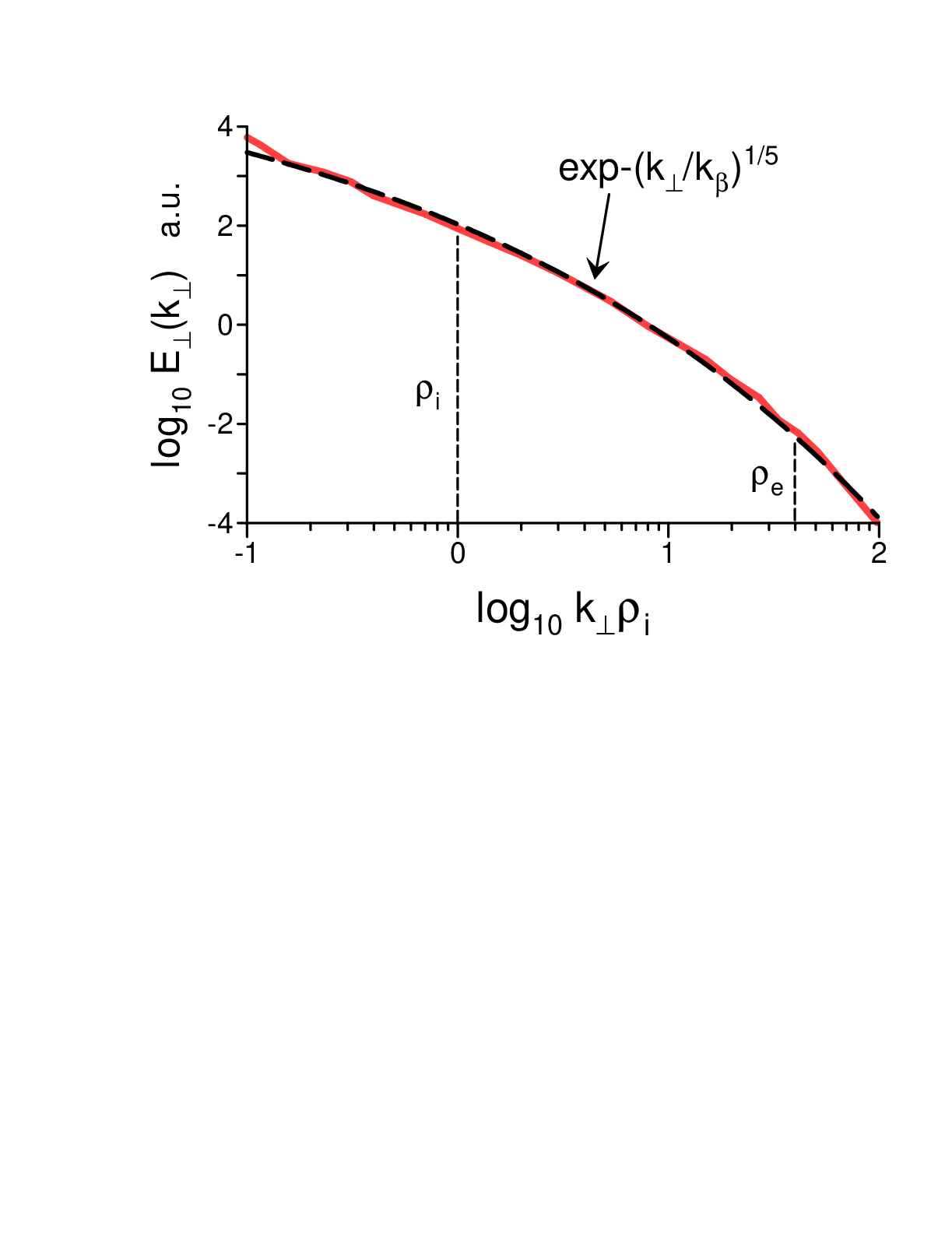} \vspace{-5.5cm}
\caption{ Perpendicular magnetic energy spectrum for a non-relativistic gyrokinetic 3D numerical simulation.}
\end{figure}
\begin{figure} \vspace{-1.2cm}\centering \hspace{-1.cm}
\epsfig{width=.46\textwidth,file=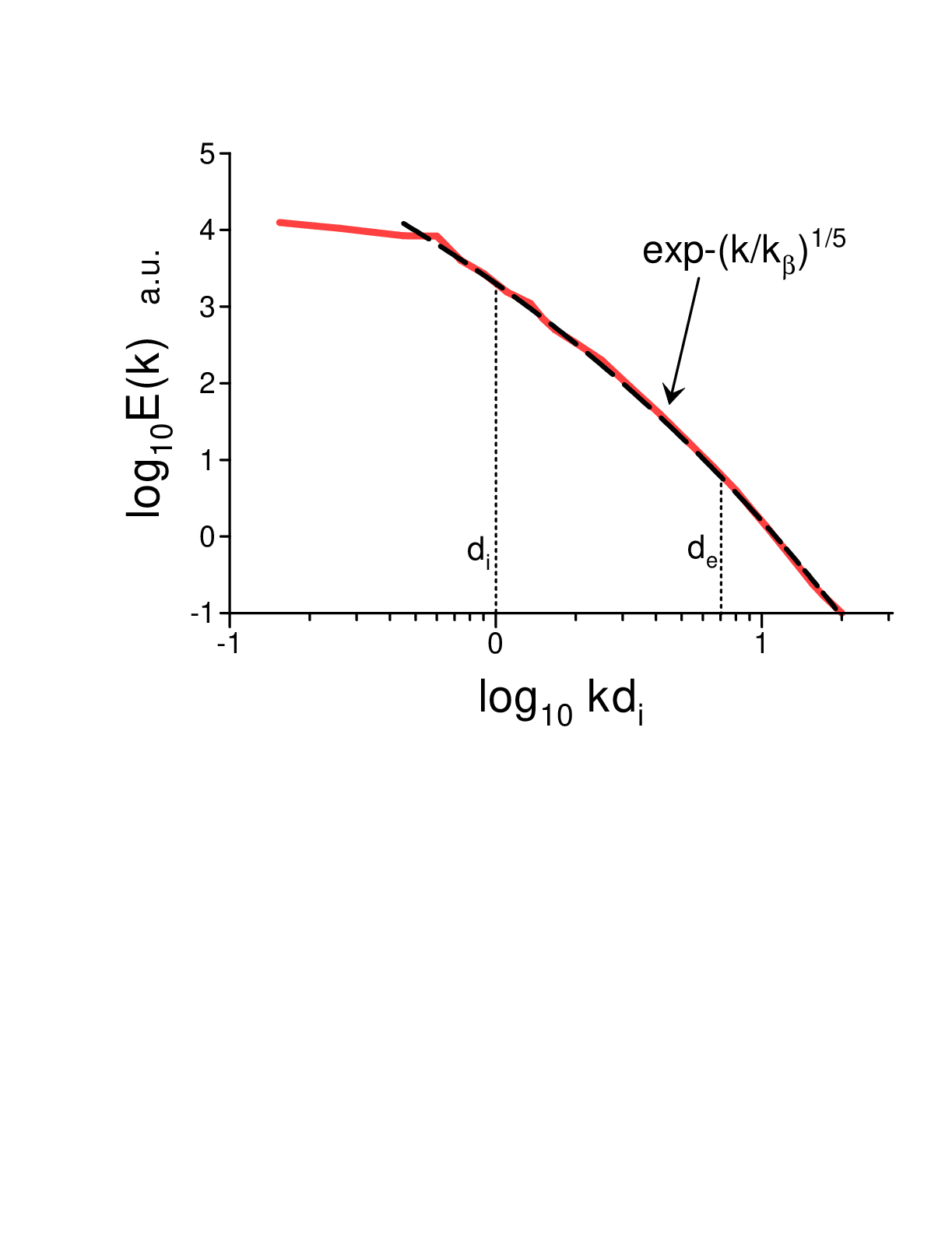} \vspace{-4.6cm}
\caption{ Omnidirectional magnetic energy spectrum for a non-relativistic fully kinetic 3D (decaying) numerical simulation.}
\end{figure}
 
   Figure 6 shows the omnidirectional magnetic energy spectrum obtained in a recent non-relativistic fully kinetic 3D simulation with an imposed uniform magnetic field ${\bf B}_0$.  The external drive was absent so the system was decaying and the spectrum was taken for $t=45 \omega_{ci}^{-1}$ (where $\omega_{ci}$ is the ion cyclotron frequency). The spectral data were taken from Fig. 4b of a recent paper \cite{yang}. The dashed curve indicates the stretched exponential spectrum Eq. (23). \\

\section{In the solar wind and  Earth's magnetosphere}

\subsection{Solar wind}

    One of the main obstacles in the studying of chaotic/turbulent fluctuations of the magnetic field at the kinetic range of scales in the solar wind is the presence of
whistler wave instabilities since the instabilities occur in this range (see, for instance, a recent paper \cite{wit} and references therein). \\

 In the same paper \cite{wit} results of measurements produced in the near-sun solar wind by the search-coil magnetometer onboard the Parker Solar Probe during its first eight solar encounters were reported.\\

  A specially chosen 2789 events with a low level of coherent wave activity (e.g. the whistler waves) were analyzed. Most of these events belong to slow Alfv\'{e}nic winds. For the different solar encounters, the range of cyclotron frequencies varies in the range of 2-10 kHz for electrons and 1–8 Hz for ions.\\ 

\begin{figure} \vspace{-0.5cm}\centering \hspace{-1.3cm}
\epsfig{width=.45\textwidth,file=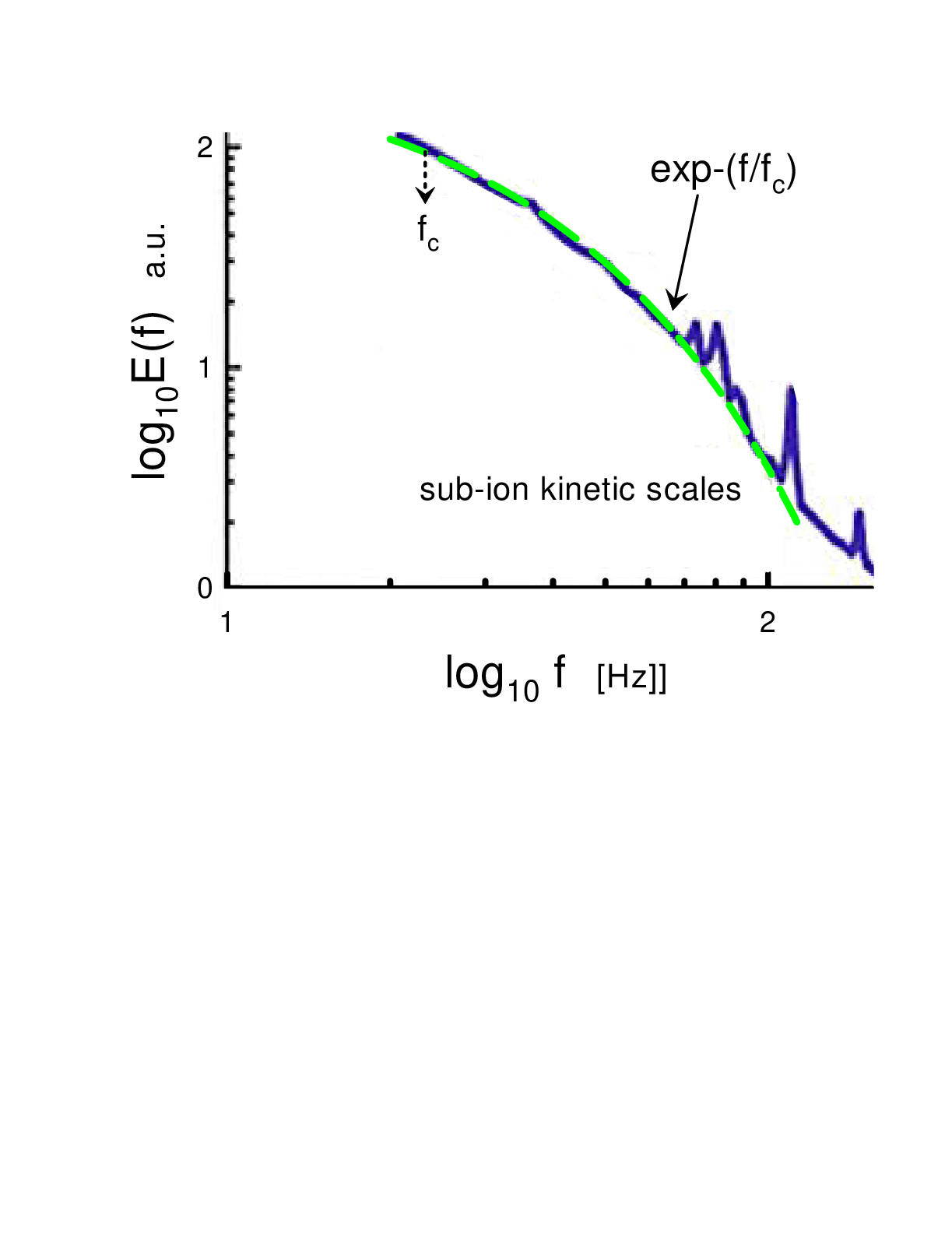} \vspace{-4.7cm}
\caption{ Power spectrum of the $B_v$ component of the magnetic field measured in the solar wind (for a set of events with a low level of coherent wave activity) at the distance from the Sun $R =0.23$ AU.}
\end{figure}
 
    Figure 7 shows the power spectrum of the $B_v$-component of the measured magnetic field (belonging to this set of events) at the distance to the Sun $R = 0.23$ AU. The magnetic field components in the instrument reference frame were denoted as $B_u$, $B_v$, and $B_w$. The  $B_u$-component is approximately antiparallel to the direction of the solar wind. The spectral data were taken from Fig. 6b of the paper \cite{wit}.  The authors of the Ref. \cite{wit} consider the frequency range shown in Fig. 7 as a sub-ion kinetic one. \\

\begin{figure} \vspace{-0.5cm}\centering 
\epsfig{width=.45\textwidth,file=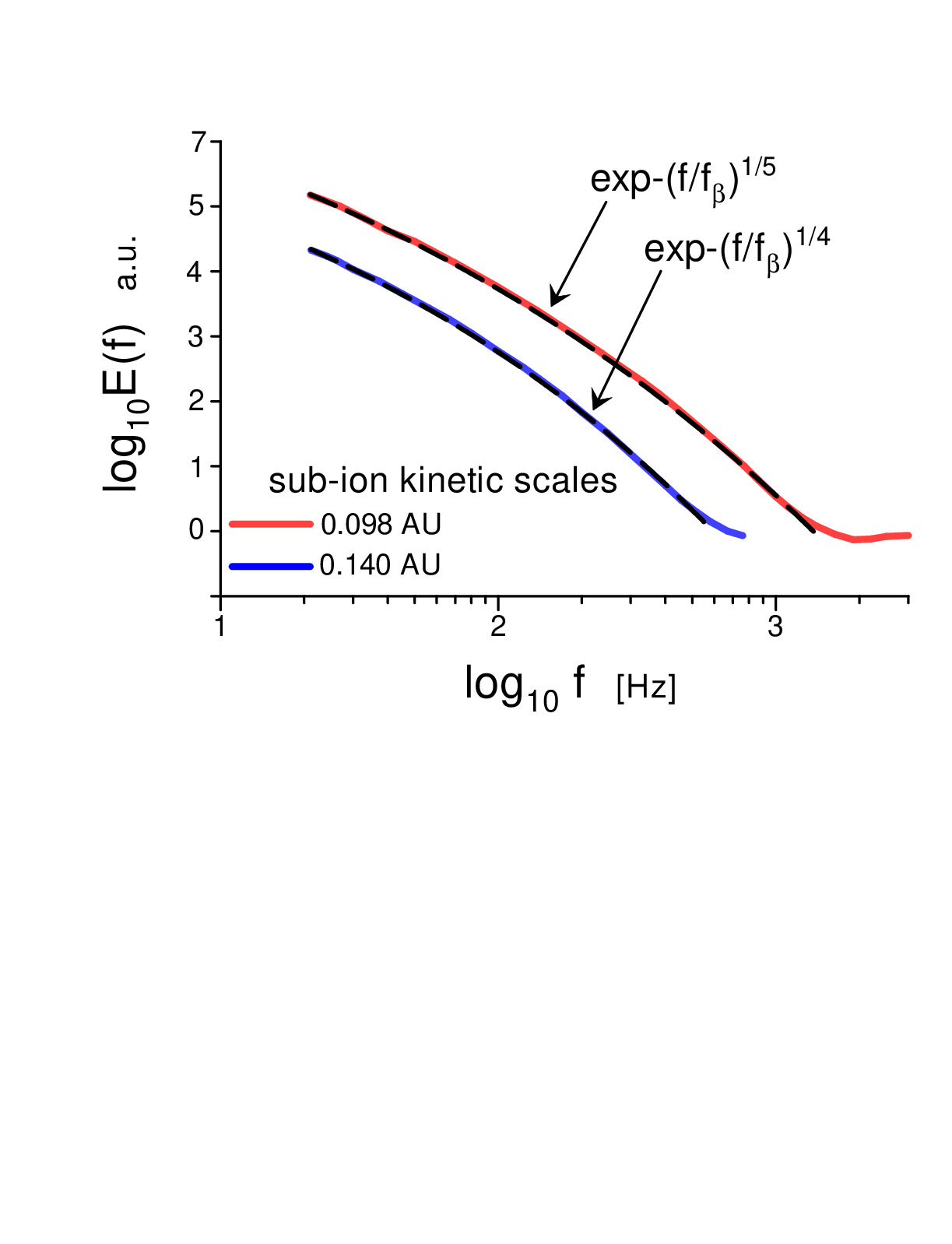} \vspace{-4.6cm}
\caption{ As in the Fig. 7 but for $R= 0.098$ and $R= 0.140$AU.}
\end{figure}
\begin{figure} \vspace{-0.6cm}\centering \hspace{-0.8cm}
\epsfig{width=.45\textwidth,file=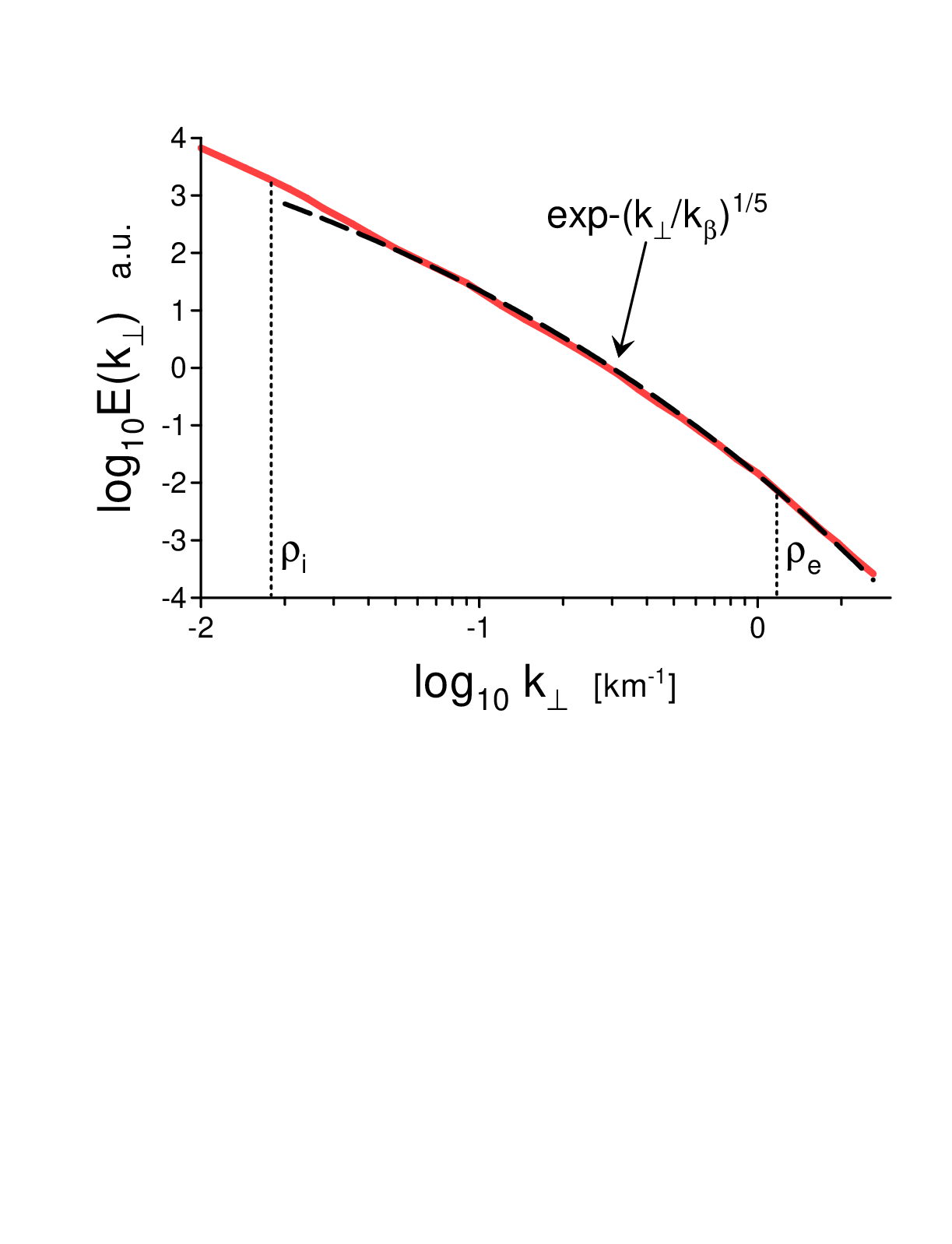} \vspace{-4.7cm}
\caption{ Spectrum of the magnetic field fluctuations measured in the solar wind by the STAFF instrument onboard the Cluster spacecraft at $R=1$ AU (for an event with a low level of coherent wave activity).}
\end{figure}

    Figure 8 shows the analogous spectrum but for $R=0.098$ and $R=0.140$ AU. The spectral data were taken from the same Fig. 6b of the paper \cite{wit}.  \\

     According to Taylor's ``frozen-in'' hypothesis, the spectra shown in Figs. 7 and 8 can have a physical interpretation as a wavenumber spectrum with the wave number $k \simeq 2\pi f/V_0$, where $V_0$ is the mean speed of the spatial magnetic structures passing the probe. This hypothesis is widely used to interpret measurements of solar wind. Application of this hypothesis to the kinetic scales is often questioned mainly due to the presence of the whistlers. Therefore, the spectra shown in Figs. 7 and 8 which correspond to the set of events with a low level of coherent wave activity are protected from this critique. \\

     The dashed curve in Fig. 7 indicates correspondence to the exponential spectral law Eq. (2) (deterministic chaos) and the dashed curves in Fig. 8 indicate correspondence to the stretched exponential spectral law Eq. (22) for $R=0.14$ AU and Eq. (23) for $R=0.098$ AU. It should be noted that preference of the Eq. (22) over the Eq. (23) for $R=0.14$ AU can indicate a diminished role of the Alfv\'{e}n waves in this case.\\
     
     In paper \cite{alex} analogous filtration of the whistlers was applied to the data obtained at 1 AU. Figure 9 shows a spectrum of the magnetic field fluctuations measured in the solar wind by the STAFF instrument onboard the Cluster spacecraft. The spectral data were taken from Fig. 6 of the Ref. \cite{alex}. Taylor's ``frozen-in'' hypothesis was applied by \cite{alex} to convert the frequencies into wavenumbers (see above). 
              
     The dashed curve in Fig. 9 indicates correspondence to the stretched exponential spectral law Eq. (23) (the magneto-inertial range for the distributed chaos in the presence of a mean magnetic field).\\

\subsection{Earth's magnetosphere}         
  
   The large variety of physical phenomena in the near-Earth plasma is caused by a complex interaction of the solar wind with the Earth's magnetic field. Therefore the existence of a universal mechanism that determined the dynamics of the magnetic field at kinetic scales in the the near-Earth plasma is rather problematic. It will be shown below that such a mechanism can be based on the magneto-inertial range approach.\\

  Figure 10 shows a magnetic field spectrum from the quasi-perpendicular bow shock obtained using data provided by the Magnetospheric Multiscale (MMS) spacecraft mission. A near-discontinuous transition from the solar wind to the bow shock is a prominent characteristic of the quasi-perpendicular shocks. The upstream solar wind was in the `slow' conditions in this case. The spectral data were taken from Fig. 3 of a recent paper \cite{pg}. The dashed curve in Fig. 10 indicates correspondence to the stretched exponential spectral law Eq. (22) (the magneto-inertial range for the distributed chaos). \\
  
Figure 11 shows a magnetic field spectrum from the quasi-parallel bow shock populated by kinetic scale reconnecting current sheets. The measurements were also produced by the Magnetospheric Multiscale spacecraft. The spectral data were taken from Fig. 1 of a paper \cite{chen}.

\begin{figure} \vspace{-1.3cm}\centering \hspace{-0.9cm}
\epsfig{width=.45\textwidth,file=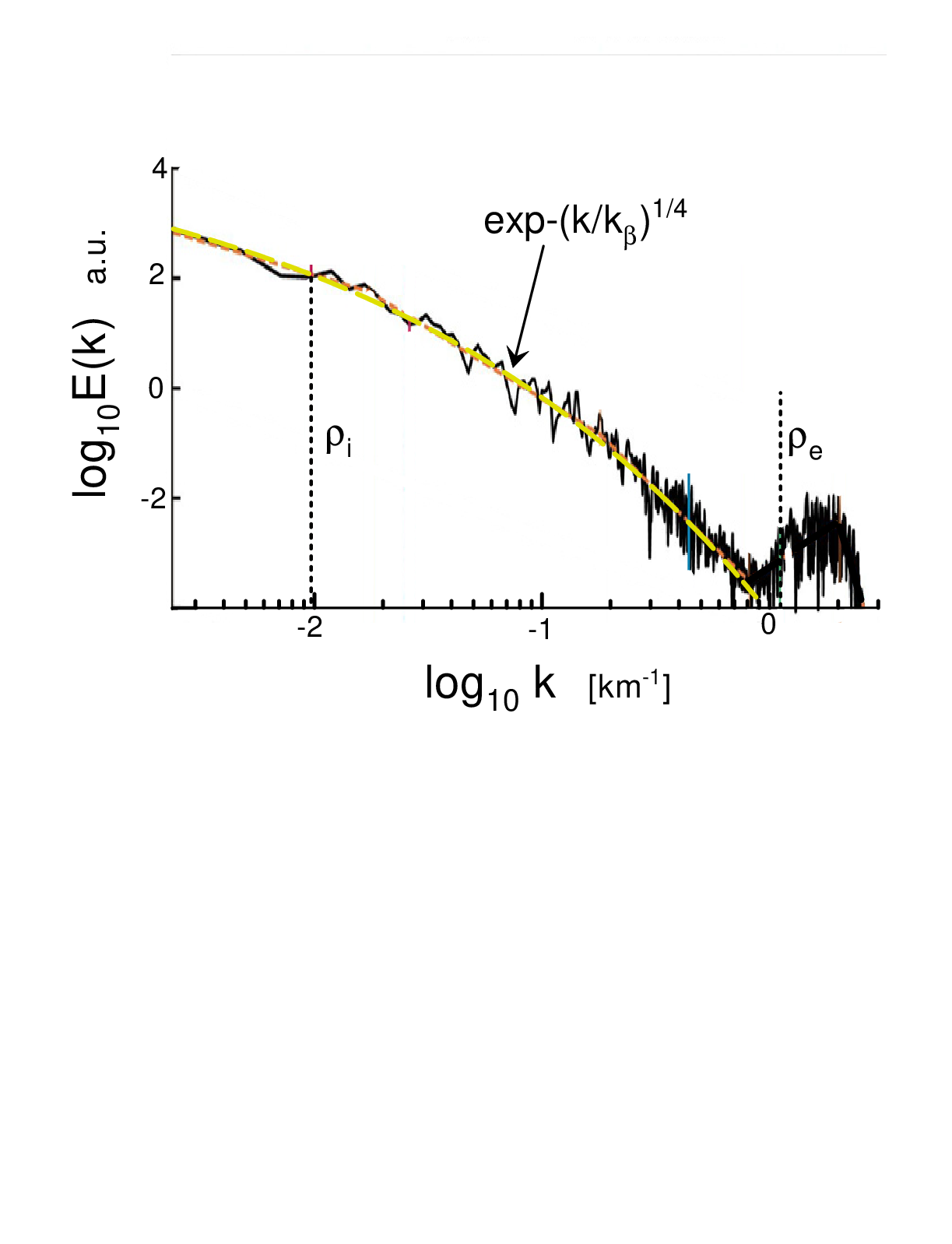} \vspace{-4.65cm}
\caption{ The trace magnetic field fluctuation spectrum measured by MMS spacecraft at the quasi-perpendicular bow shock. }
\end{figure}
\begin{figure} \vspace{-0.3cm}\centering \hspace{-1.2cm}
\epsfig{width=.47\textwidth,file=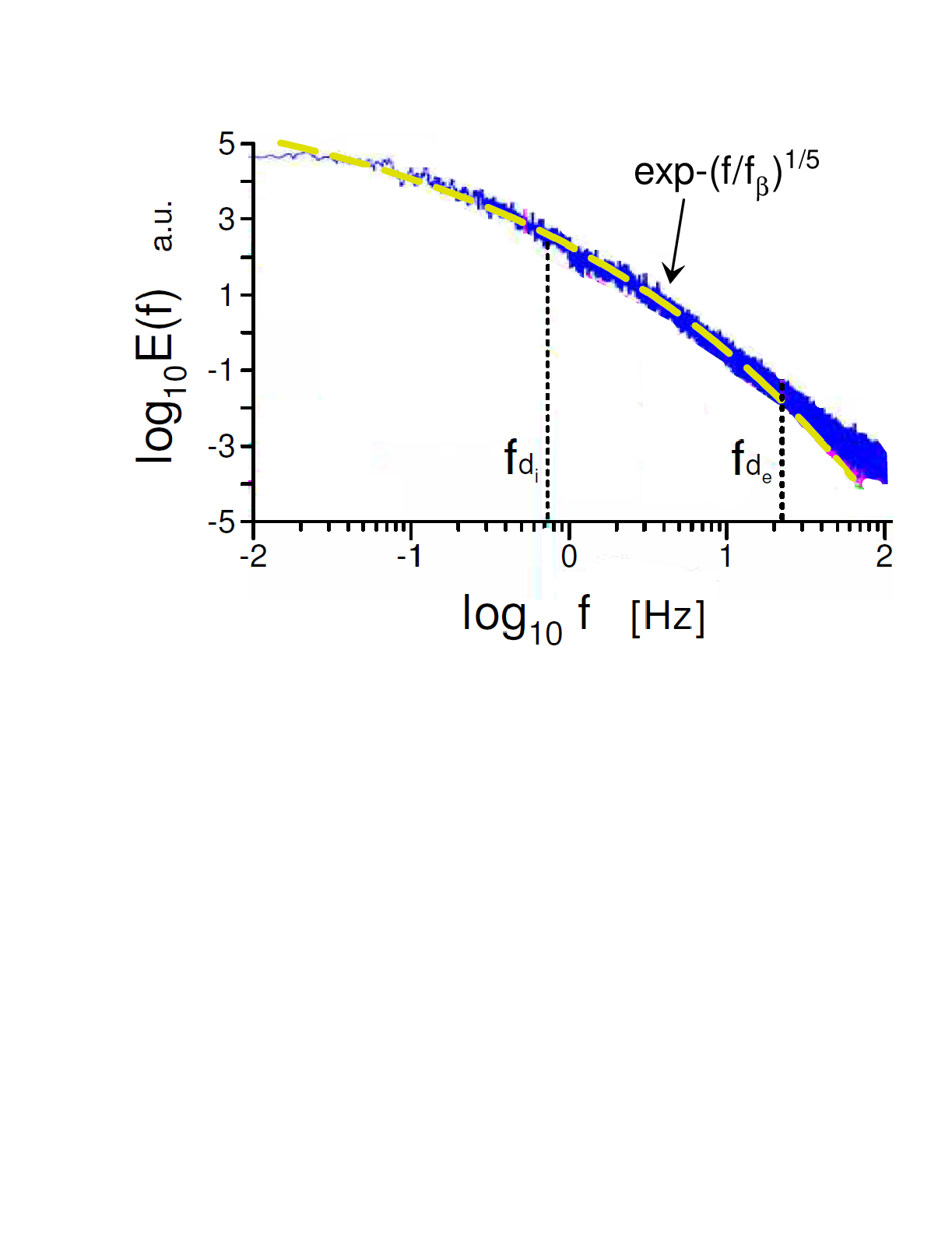} \vspace{-5.45cm}
\caption{ The trace magnetic field fluctuation spectrum measured by MMS spacecraft at the quasi-parallel bow shock populated by reconnecting current sheets. }
\end{figure}
\begin{figure} \vspace{-0.9cm}\centering \hspace{-1.2cm}
\epsfig{width=.46\textwidth,file=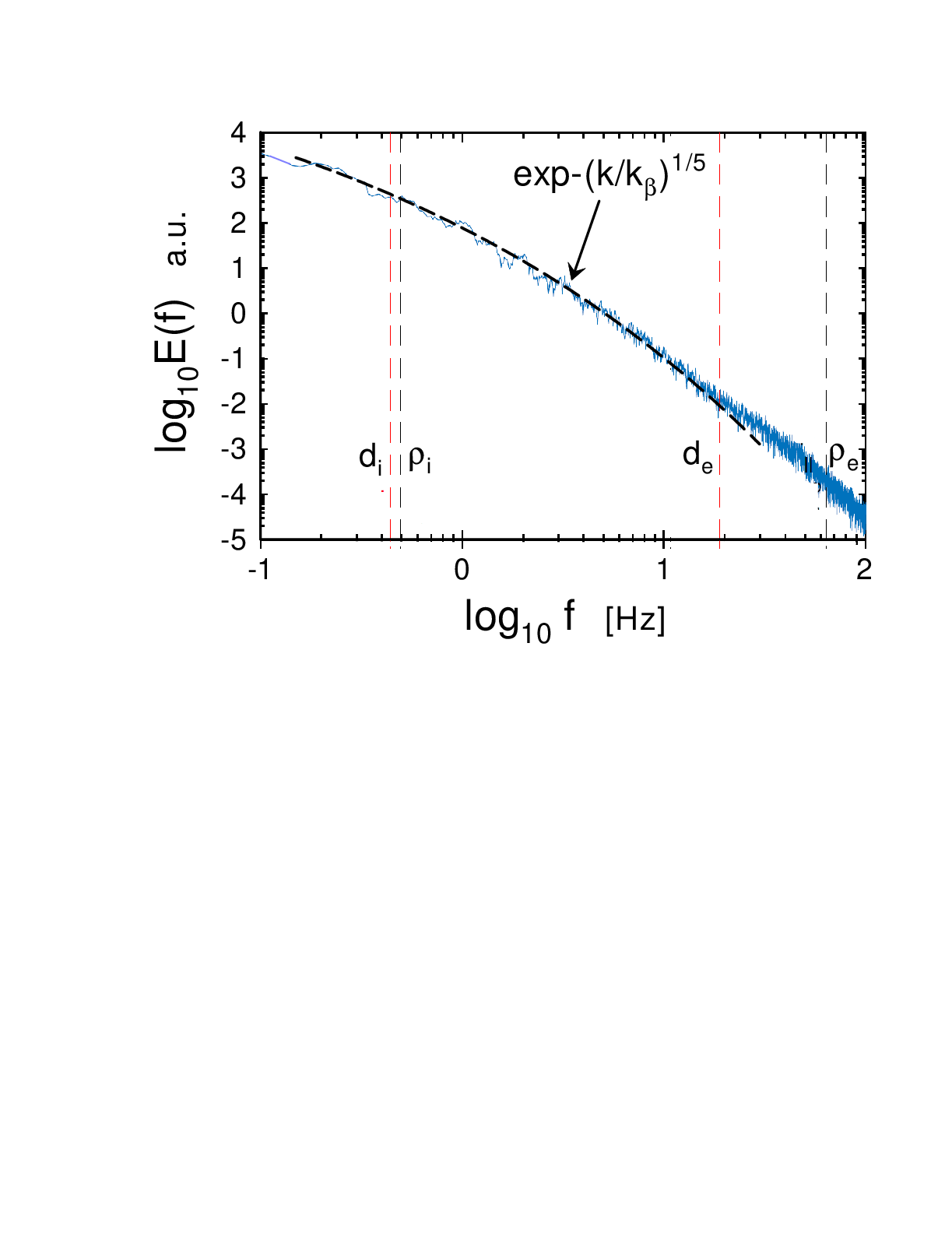} \vspace{-5.3cm}
\caption{  The trace magnetic field fluctuation spectrum measured by MMS spacecraft at the Earth’s magnetosheath close to the dusk
side magnetopause. }
\end{figure}
\begin{figure} \vspace{-0.6cm}\centering \hspace{-1.2cm}
\epsfig{width=.46\textwidth,file=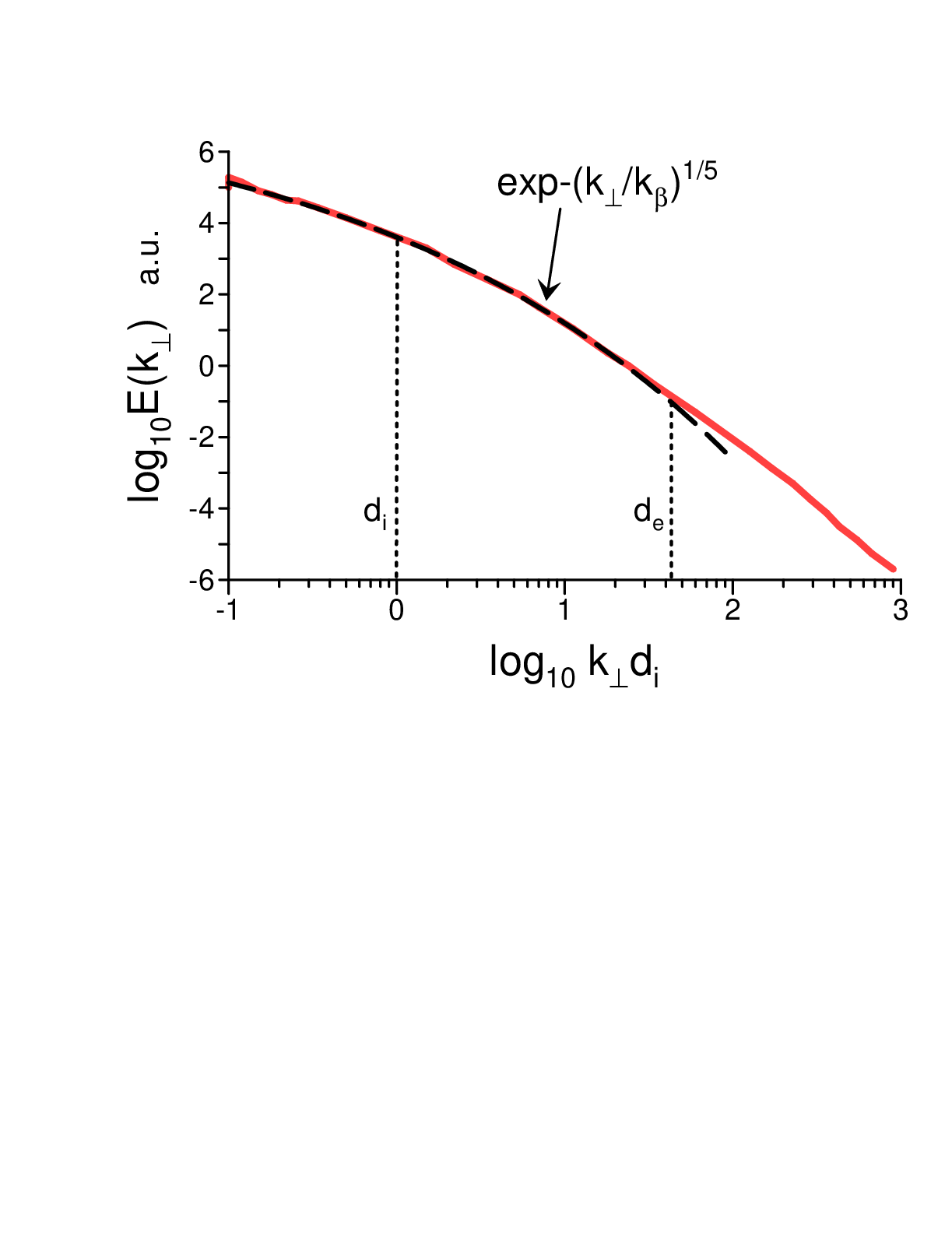} \vspace{-5cm}
\caption{ The trace magnetic field fluctuation spectrum measured by MMS spacecraft on the duskside of the Earth’s magnetopause.}
\end{figure}
  
  The dashed curve in Fig. 11 indicates correspondence to the stretched exponential spectral law Eq. (23) (the magneto-inertial range for the distributed chaos in the presence of a mean magnetic field). 

   The helical distributed chaos is convected from the bow shock into the magnetosheath.\\

  Figure 12 shows the trace magnetic field fluctuation spectrum measured by Magnetospheric Multiscale spacecraft at the Earth’s magnetosheath (close to the dusk
side magnetopause) during a period with a low activity of mirror modes and whistler waves (see for a more detailed description of the conditions Ref. \cite{cb}). The spectral data were taken from Fig. 5 of Ref. \cite{ckh}. The dashed curve in Fig. 12 indicates correspondence to the stretched exponential spectral law Eq. (23) (the magneto-inertial range for the distributed chaos in the presence of a mean magnetic field).\\

  Figure 13 shows the trace magnetic field fluctuation spectrum measured by Magnetospheric Multiscale spacecraft on the duskside of the Earth’s magnetopause during a long period when a continuous train of Kelvin–Helmholtz waves was observed. The spectral data were taken from Fig. 2a of Ref. \cite{franci}. The transformation from frequency to the wavenumber perpendicular to the ambient magnetic field was made in Ref. \cite{franci} using the Taylor's `frozen-in' hypothesis (see above). The dashed curve in Fig. 13 indicates correspondence to the stretched exponential spectral law Eq. (23) (the magneto-inertial range for the distributed chaos in the presence of a mean magnetic field). \\

\begin{figure} \vspace{-0.6cm}\centering \hspace{-1.2cm}
\epsfig{width=.43\textwidth,file=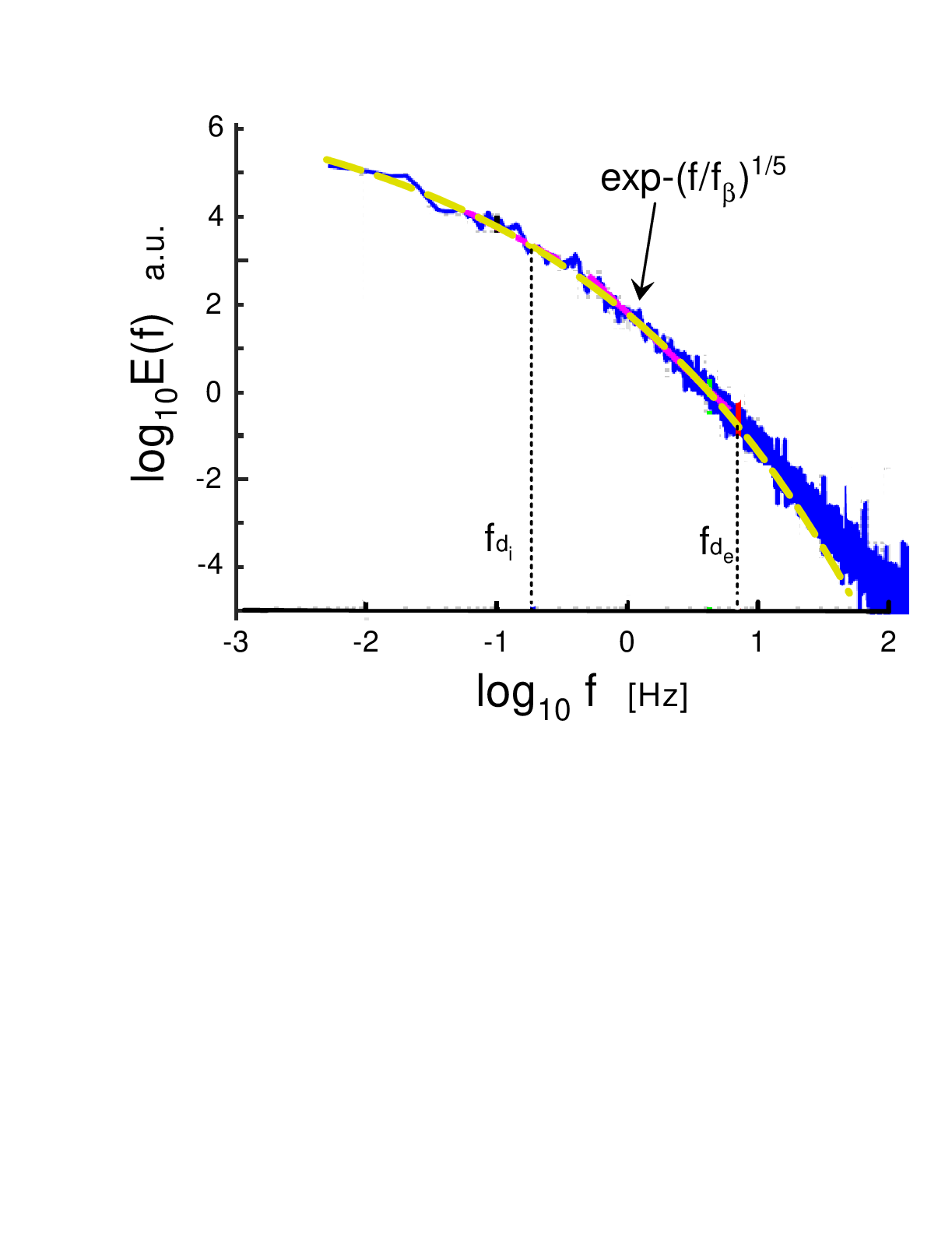} \vspace{-4.4cm}
\caption{ Power spectrum of a magnetic field component directly related to a transient reconnection event at the Earth's magnetotail (the measurements were produced by the MMS spacecraft).}
\end{figure}
\begin{figure} \vspace{-0.4cm}\centering \hspace{-1.2cm}
\epsfig{width=.45\textwidth,file=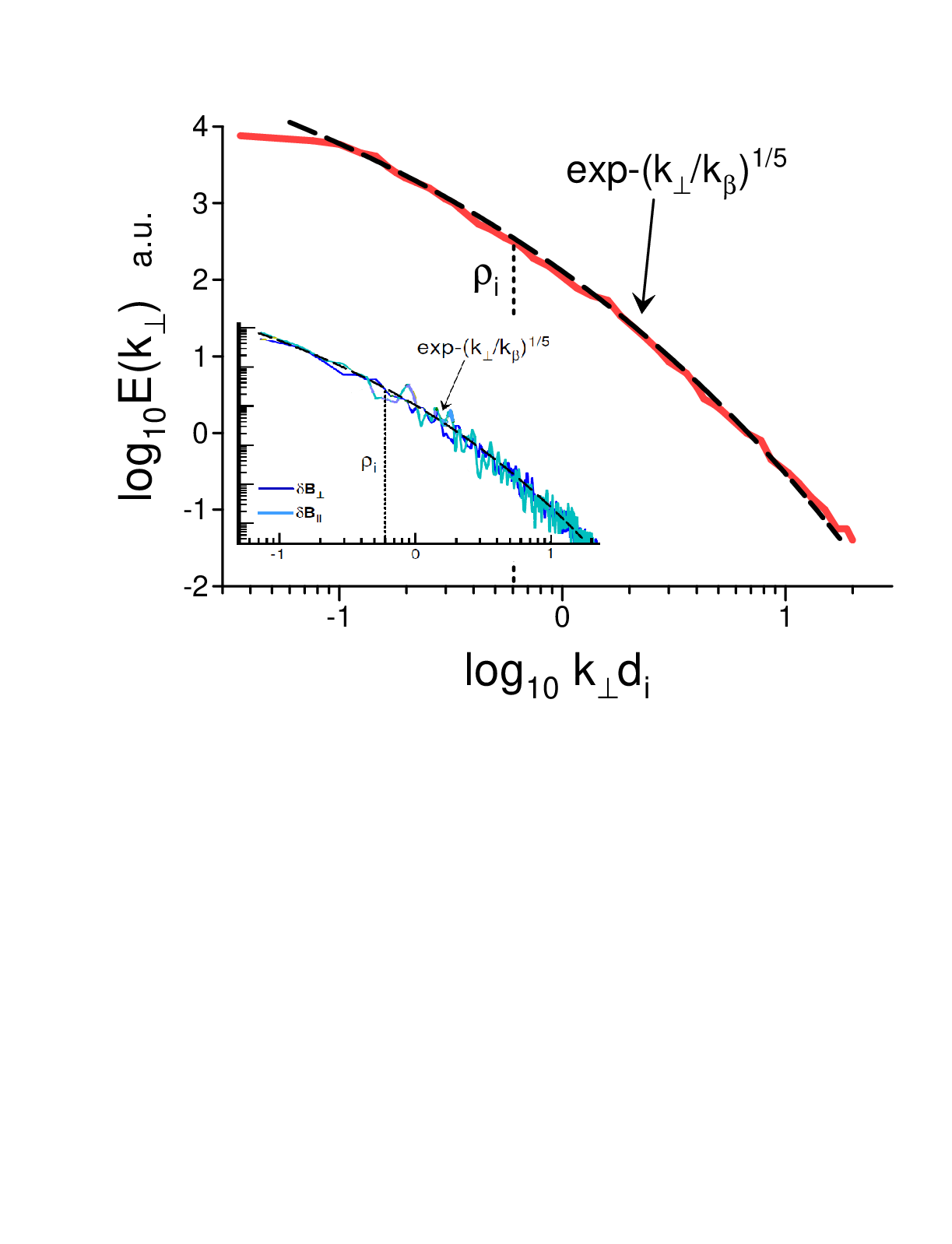} \vspace{-4.8cm}
\caption{ Ensemble-averaged trace magnetic field spectrum measured by MMS spacecraft in 24 magnetic reconnection jets at the plasma sheet of the magnetotail. The insert shows power spectra of the parallel and perpendicular components of the magnetic field fluctuations measured in one of these reconnection jets as an example. }
\end{figure}
\begin{figure} \vspace{-0.4cm}\centering \hspace{-1cm}
\epsfig{width=.51\textwidth,file=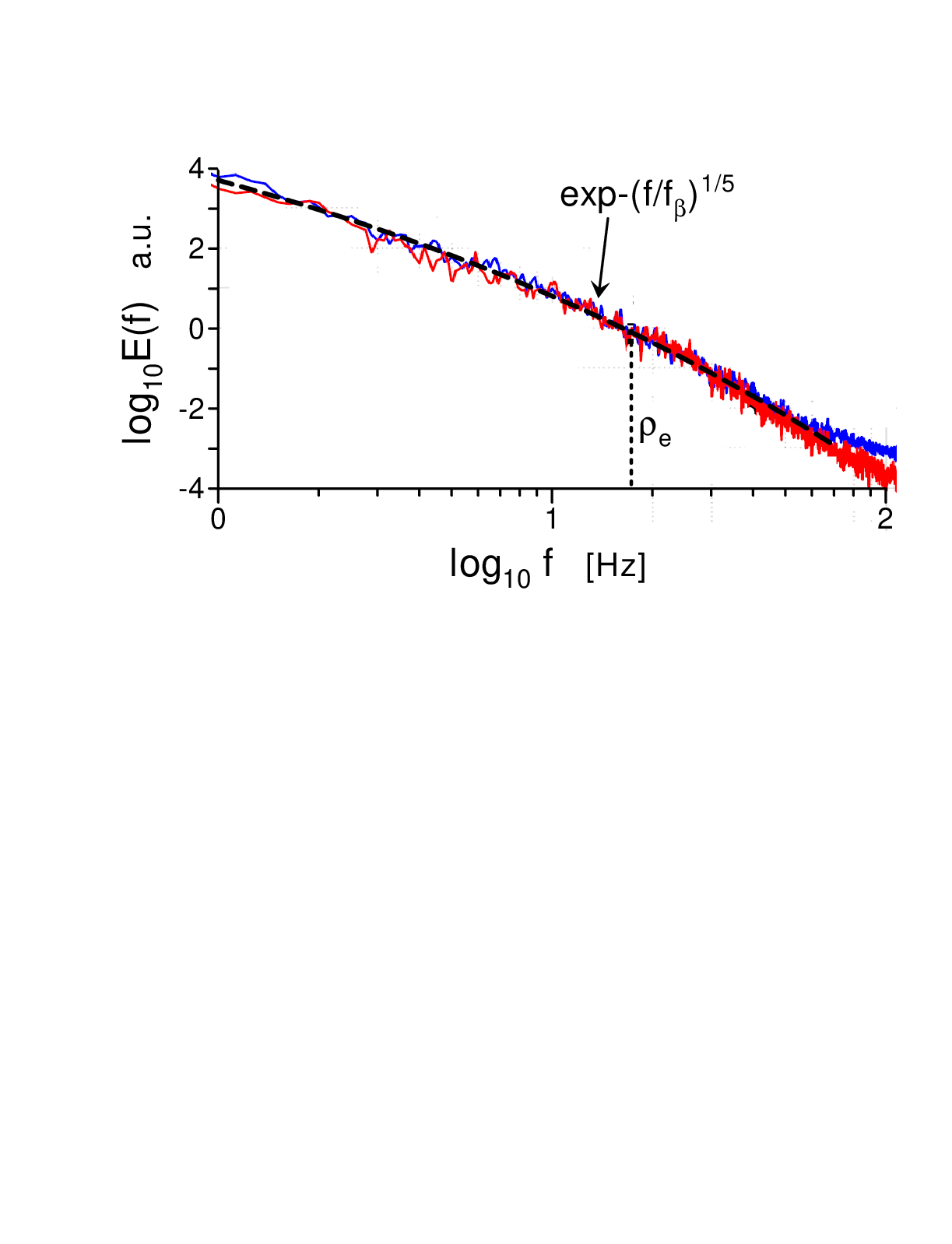} \vspace{-6.5cm}
\caption{ Power spectra of magnetic field fluctuations perpendicular (blue) and parallel (red) to the mean magnetic field computed for the data obtained by the MMS in the quasi-parallel bow shock with a weak presence of the whistlers at the kinetic scales. }
\end{figure}
 
  Figure 14 shows the power spectrum of a magnetic field component directly related to a transient reconnection event at the Earth's magnetotail far away from the magnetopause boundary layer. The measurements were produced by the Magnetospheric Multiscale spacecraft. The spectral data were taken from Fig. 5c of a recent paper \cite{jin}. The dashed curve in Fig. 14 indicates correspondence to the stretched exponential spectral law Eq. (23) (the magneto-inertial range for the distributed chaos in the presence of a mean magnetic field). \\

   Figure 15 shows the ensemble-averaged trace magnetic field spectrum measured by Magnetospheric Multiscale spacecraft in magnetic reconnection jets at the plasma sheet of the magnetotail. The ensemble consists of the 24 reconnection jets which show signatures of fully developed chaos/turbulence (the spectral data were taken from Fig 2a of a recent paper \cite{rich}). The insert shows power spectra of the parallel and perpendicular components of the magnetic field fluctuations measured in one of these reconnection jets as an example (the spectral data were taken from Fig 1d of the paper \cite{rich}). The dashed curves in Fig. 15 indicate correspondence to the stretched exponential spectral law Eq. (23) (the magneto-inertial range for the distributed chaos in the presence of a mean magnetic field).

\subsection{Sub-electron scales}  
   
   One can see from the examples that for the magnetospheric plasmas the approach to the kinetic scales, based on the distributed chaos notion, is apparently restricted by the sub-ionic scales. High-quality measurements at the {\it electron} scales are still rare and just at these scales the coherent wave activity (like whistlers) is typically the most significant (the whistlers violate Taylor's ``frozen-in'' hypothesis and can corrupt the chaotic/turbulent nature of the magnetic field fluctuations, see above). \\

   In a paper \cite{bre} the magnetic field spectra were computed from the data obtained by the Magnetospheric Multiscale mission in the Earth's magnetosheath for the quasi-parallel and quasi-perpendicular bow shocks. For the data obtained in the quasi-perpendicular bow shock intense whistlers at the kinetic scales were observed whereas in the quasi-parallel bow shock the presence of the whistlers at the kinetic scales was weak. \\
   
   Figure 16 shows the power spectra of the components of the magnetic field fluctuations perpendicular (blue) and parallel (red) to the mean magnetic field reported in the Ref. \cite{bre} for the quasi-parallel bow shock (the spectral data were taken from Fig. 3 of the Ref. \cite{bre}). The dashed curve in Fig. 16 indicates correspondence to the stretched exponential spectral law Eq. (23) (the magneto-inertial range for the distributed chaos in the presence of a mean magnetic field).\\

\begin{figure} \vspace{-0.1cm}\centering \hspace{-1.2cm}
\epsfig{width=.45\textwidth,file=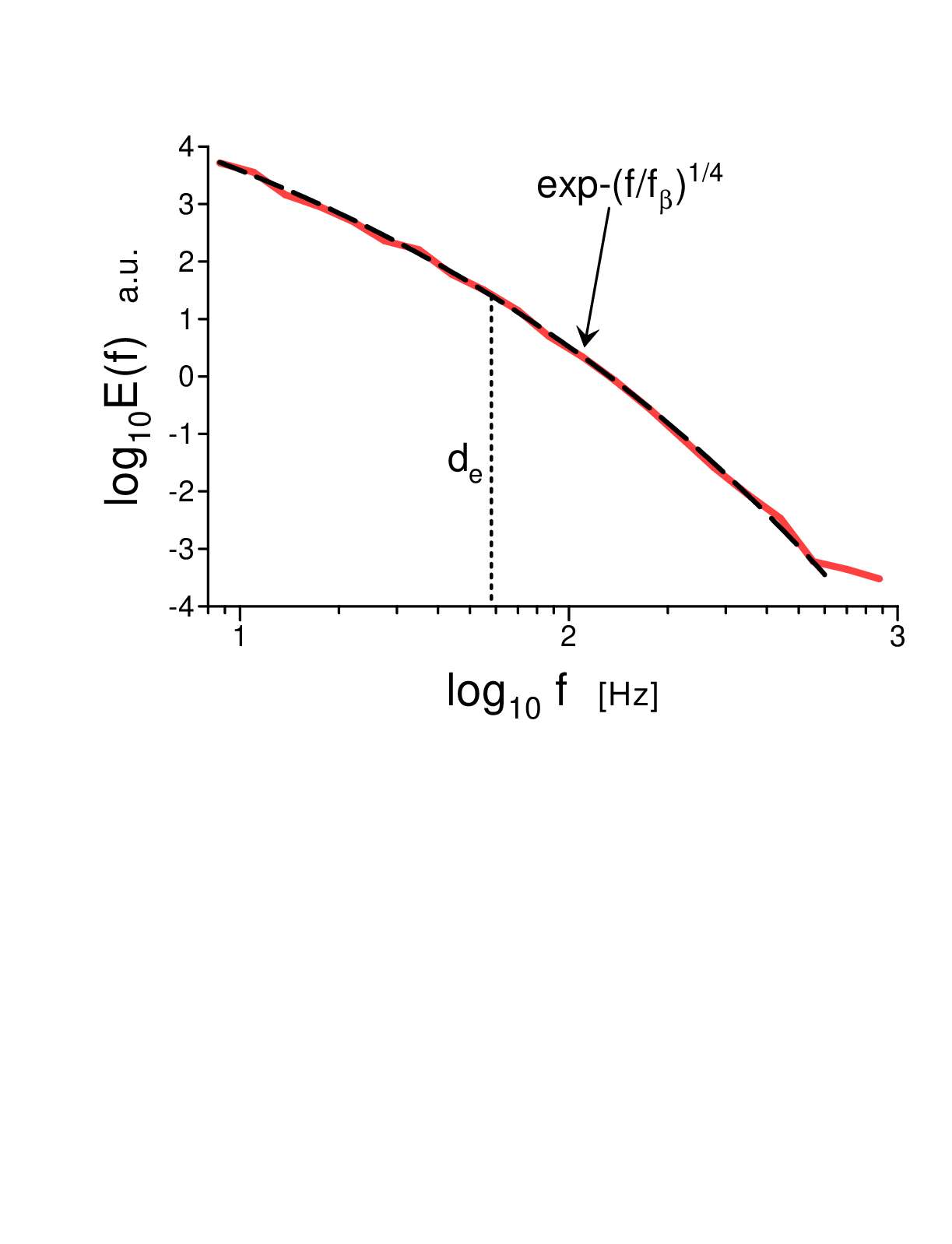} \vspace{-4.5cm}
\caption{ The trace magnetic field fluctuation spectrum computed using {\it Cluster} data obtained in the Earth's magnetosheath (with filtered-out whistlers). }
\end{figure}
   
\begin{figure} \vspace{-0.5cm}\centering \hspace{-1.2cm}
\epsfig{width=.45\textwidth,file=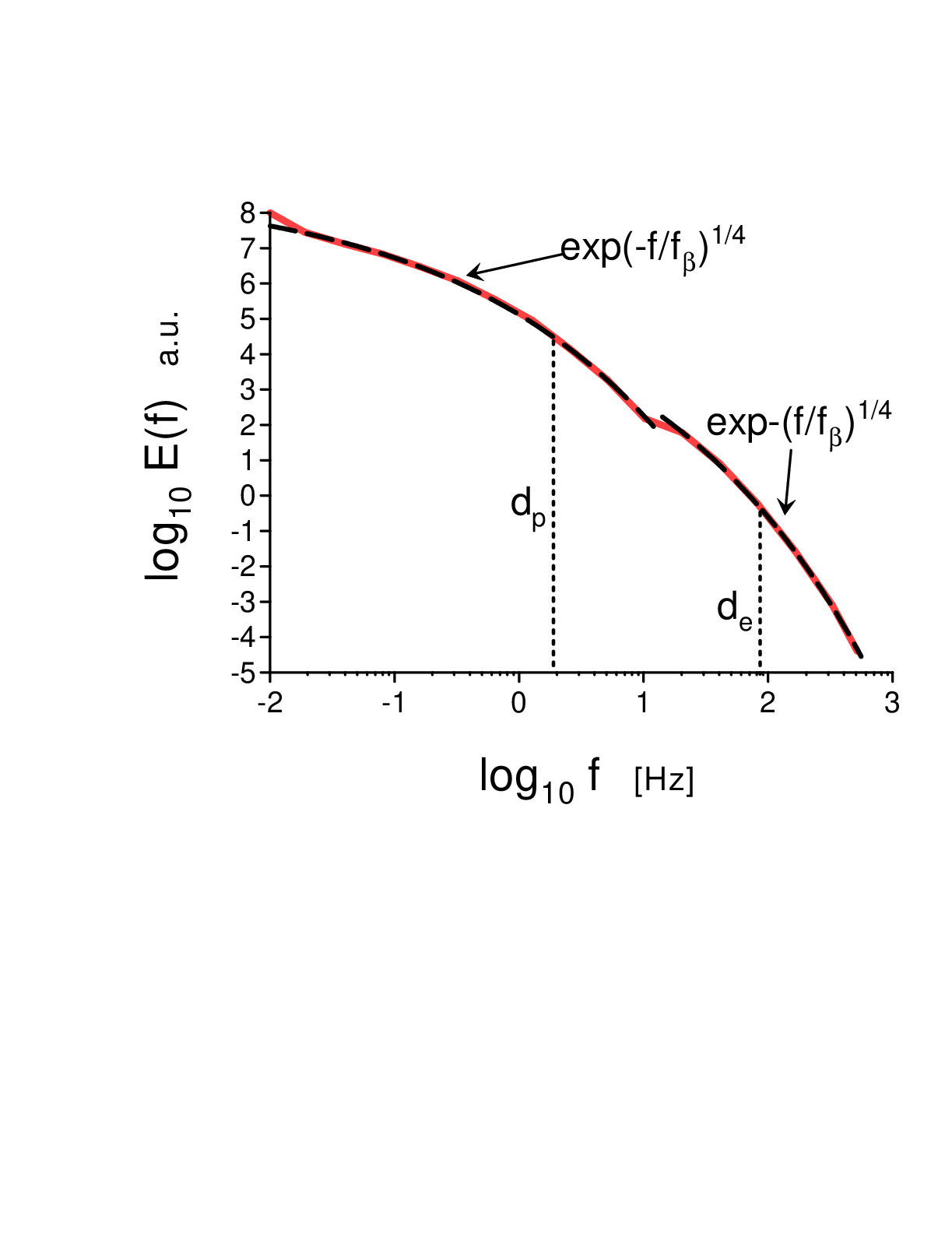} \vspace{-3.8cm}
\caption{ The trace magnetic field fluctuation spectrum computed using {\it Cluster} data obtained in the Earth's magnetosheath with high plasma's beta. }
\end{figure}
   
   In paper \cite{mat} the data obtained with {\it Cluster} mission in the Earth's magnetosheath were analyzed with special attention to the problem caused by whistlers. Figure 17 shows the trace magnetic field spectrum obtained in the magnetosheath with filtered-out whistlers. The spectral data were taken from Fig. 1 of the paper \cite{mat}. The dashed curve in Fig. 17 indicates correspondence to the stretched exponential spectral law Eq. (22) (the magneto-inertial range for the distributed chaos). \\
   
   Figure 18 shows the trace magnetic field spectrum obtained with the {\it Cluster} mission in the Earth's magnetosheath and corresponding to high plasma's beta. The spectral data were taken from Fig. 4a of the paper \cite{mat}. The dashed curves in Fig. 18 indicate correspondence to the stretched exponential spectral law Eq. (22) (the magneto-inertial range for the distributed chaos). The vertical dotted lines indicate the positions of the proton $d_p$ and electron $d_e$ inertial length respectively. One can see that in this case there are two magneto-inertial ranges: one centered on the $d_p$ and another centered on $d_e$. The former range is overlapped with the MHD scales. \\

\section{Conclusions and Discussion}
  
  The above-given examples show that magnetic helicity (as an adiabatic invariant) can control the chaotic/turbulent magnetic field dynamics at kinetic scales of the collisionless plasmas (directly or through the Kolmogorov-Iroshnikov phenomenology applied to the characteristic values of magnetic field and wavenumber).\\
  
   This phenomenon takes place even in the cases of zero (or negligible) global helicity due to the spontaneous breaking of the local reflectional symmetry typical for chaotic/turbulent dynamics.\\
   
  The problem of the applicability of the above-suggested approach to the sub-electron scales of the space plasmas is still open (see Section 5c). Since for the kinetic numerical simulations applicability of this approach to the sub-electron scales can also be observed this question needs additional investigation. \\
  
  It should be noted in this respect that in a recent paper \cite{mil} a model considering the kinetic effects related to the electron inertia and characterized by the electron skin-depth $d_e$ was suggested. This model has two quadratic invariants: the total energy and a generalized cross-helicity (see a recent paper \cite{ber3} for an application of the distributed chaos notion to the chaotic/turbulent dynamics of the magnetic field dominated by cross-helicity). \\

    The interplay of the coherent wave activity (mirror modes, whistler wave instabilities, and so on) and of the chaotic/turbulent dynamics at the same kinetic scales is a rather important factor that can complicate the observed picture. Therefore, the above-discussed filtration of the coherent wave activity can be very useful in separating these phenomena in space plasmas, that allows a more detailed study of the chaotic/turbulent plasma dynamics at the kinetic scales. \\
    
    Despite considerable differences in the scales and physical parameters results of numerical simulations are in quantitative agreement with the measurements in the space plasmas.  

\section{Acknowledgments }

  I thank H.K. Moffatt, A. Pikovsky, and J.V. Shebalin for stimulating discussions.
  
\clearpage  


\begin{thebibliography}{}
\bibitem{fm} U. Frisch and R. Morf,  Phys. Rev. {\bf 23}, 2673 (1981).
\bibitem{swinney1} A. Brandstater and H. L. Swinney, Phys. Rev. A {\bf 35}, 2207 (1987).
\bibitem{mm1} J. E. Maggs and G. J. Morales, Phys. Rev. Lett.  {\bf 107}, 185003 (2011). 
\bibitem{mm2} J.E. Maggs and G.J. Morales, Phys. Rev. E {\bf 86}, 015401(R) (2012)
\bibitem{mm3} J. E. Maggs and G. J. Morales, Plasma Phys. Control. Fusion {\bf 54}, 124041 (2012).
\bibitem{kds} S. Khurshid, D. A. Donzis, and K. R. Sreenivasan, Phys. Rev. Fluids {\bf 3}, 082601(R) (2018).
\bibitem{wu} X.-Z. Wu, L. P. Kadanoff, A. Libchaber and M. Sano, Phys. Rev. Lett. {\bf 64}, 2140 (1990).
\bibitem{map} P. D. Mininni, A. Alexakis, A. Pouquet, J. Plasma Phys. {\bf 73}, 377 (2007).
\bibitem{ber4} A Bershadskii, Fundamental Plasma Physics {\bf 11}, 100066 (2024).
\bibitem{mt} H. K. Moffatt and A. Tsinober, Annu. Rev. Fluid Mech. {\bf 24}, (1992) 281 (1992).
\bibitem{pm} J. M. Polygiannakis and X. Moussas, Plasma Phys. Control. Fusion {\bf 43}, (2001) 195 (2001).
\bibitem{sch1} A. A. Schekochihin, S. C. Cowley, W. Dorland, G. W.Hammett, G. G.Howes, E. Quataert and T. Tatsuno, Astrophys. J. Suppl. Ser.  {\bf 182}, (2009) 310 (2009).
\bibitem{mg} W. H. Matthaeus and M. L. Goldstein, J. Geophys. Res. {\bf 87}, 6011 (1982).
\bibitem{shebalin} J. V. Shebalin, Phys. Plasmas {\bf 20}, (2013) 102305 (2013).
\bibitem{my} A. S. Monin, and A. M. Yaglom, Statistical Fluid Mechanics, Vol. II: Mechanics of Turbulence (Dover Pub. NY, 2007).
\bibitem{jkb} N. L. Johnson, S. Kotz, and N. Balakrishnan, Continuous Univariate Distributions, Vol. 1, (Wiley NY, 1994)
\bibitem{moff1} H. K. Moffatt, J. Fluid Mech. {\bf 35}, (1969) 117 (1969).
\bibitem{moff2} H. K. Moffatt, J . Fluid Mech. {\bf 159}, (1985) 359 (1985).
\bibitem{kerr} R. M. Kerr,  In: Elementary Vortices and Coherent Structures, Proceedings of the IUTAM Symposium Kyoto, 1-8 (2004).
\bibitem{hk} D. D. Holm, R. M. Kerr, Physics of Fluid, {\bf 19}, 025101 (2007).
\bibitem{lt} E. Levich and A. Tsinober, Phys. Lett. A {\bf 93}, 293 (1983).
\bibitem{cs} L. Comisso and L. Sironi, ApJ Lett. {\bf 936}, L27 (2022).
\bibitem{bs1} A. Bershadskii, and K. R. Sreenivasan, Phys. Rev. Lett. {\bf 93}, (2004) 064501 (2004).
\bibitem{ir} R. S. Iroshnikov, Astronomicheskii Zhurnal, {\bf 40}, 742 (1963) (English translation in Soviet Astronomy {\bf 7}, 566 (1964)).
\bibitem{jon} D. C. Johnston, Phys. Rev. B {\bf 74}, 184430 (2006).
\bibitem{zd} V. Zhdankin, ApJ {\bf 922}, 172 (2021).
\bibitem{wan} M. Wan, W. H. Matthaeus, V. Roytershteyn, H. Karimabadi, T. Parashar, P. Wu, and M. Shay, Phys. Rev. Lett. {\bf 114}, 175002 (2015).
\bibitem{thd} J. M. TenBarge, G. G. Howes, and W. Dorland, ApJ {\bf 774}, 139 (2013).
\bibitem{th} J. M. TenBarge, G. G. Howes, ApJL  {\bf 771}, L27 (2013).
\bibitem{yang} Y. Yang, W. H. Matthaeus, S. Roy, V. Roytershteyn, T. N. Parashar, Ri. Bandyopadhyay, and M. Wan, ApJ {\bf 929}, 142 (2022).
\bibitem{wit} T. Dudok de Wit, V. V. Krasnoselskikh, O. Agapitov, C. Fromen, A. Larosa, S. D. Bale, T. Bowen, K. Goetz, P. Harvey, G. Jannet, et al., Journal of Geophysical Research (Space Physics) {\bf 127}, e2021JA030018 (2022).
\bibitem{alex} O. Alexandrova, C. Lacombe, A. Mangeney, R. Grappin, and M. Maksimovic, ApJ {\bf 760}, 121 (2012).
\bibitem{pg} J. Plank and I. L. Gingell, Phys. Plasmas {\bf 3}, 082906 (2023).
\bibitem{chen} L. J. Chen, N. Bessho, J. A. Bookbinder, D. Caprioli, M. Goldstein, H. Ji, L. K. Jian, H. Karimabadi, Y. Khotyaintsev, K. G. Klein, et al., arXiv:1908.04192 (2019).
\bibitem{cb} C. H. K. Chen, S. Boldyrev, ApJ {\bf 842}, 122 (2017).
\bibitem{ckh} C. H. K. Chen, K. G. Klein, and G. G. Howes, Nature Communications {\bf 10}, 740 (2019).
\bibitem{franci} L. Franci, J. E. Stawarz, E. Papini, P. Hellinger, T. Nakamura, D. Burgess, S Landi, . Verdini, L. Matteini, R. Ergun, ApJ {\bf 898}, 175 (2020).
\bibitem{jin} R. Jin, M. Zhou, Y. Pang, X. Deng, and Y. Yi, ApJ {\bf 925}, 17 (2022).
\bibitem{rich} L. Richard, L. Sorriso-Valvo, E. Yordanova, D. B. Graham, and Y. V. Khotyaintsev, Phys. Rev. Lett. {\bf 132}, 105201 (2024).
\bibitem{bre} H. Breuillard, L. Matteini, M. R. Argall, F. Sahraoui,  M. Andriopoulou, O. Le Contel, A. Retino, L. Mirioni, S. Y. Huang, D. J. Gershman, ApJ {\bf 859}, 127 (2018).
\bibitem{mat} L. Matteini, O. Alexandrova, H. K. Chen, and C. Lacombe, MNRAS {\bf 466}, 945 (2017).
\bibitem{mil} L. M. Milanese, N. F. Loureiro, M. Daschner and S. Boldyrev, Phys. Rev. Lett. {\bf 125}, 265101 (2020).
\bibitem{ber3} A. Bershadskii, Res. Notes AAS {\bf 4}, 10 (2020).
\end{thebibliography}
\end{document}